\documentclass[aps,pra,groupedaddress,
amsmath,amssymb,
multicolumn,
reprint,
floatfix,
noeprint,
noshowpacs,
a4paper,
noshowkeys,
twoside
]{revtex4-1}

\usepackage{graphicx}
\usepackage{amsmath, amssymb}
\usepackage[latin1]{inputenc}
\usepackage[T1]{fontenc}
\usepackage{color}
\usepackage{hyperref}
\hypersetup{
   bookmarks=false,
   colorlinks,
   citecolor=blue,
   filecolor=blue,
   linkcolor=black,
   urlcolor=blue
}
\usepackage{array}
\newcolumntype{L}[1]{>{\raggedright\let\newline\\\arraybackslash\hspace{0pt}}m{#1}}
\newcolumntype{C}[1]{>{\centering\let\newline\\\arraybackslash\hspace{0pt}}m{#1}}
\newcolumntype{R}[1]{>{\raggedleft\let\newline\\\arraybackslash\hspace{0pt}}m{#1}}

\makeatletter
  \renewcommand\p@figure{Fig.~}
  \renewcommand{\eqref}[1]{\textup{Eq.~\tagform@{\ref{#1}}}} % redefine \eqref to prepend ``Eq.''.
\makeatother
\newcommand{\abs}[1]{\left\vert #1 \right\vert}
\newcommand{\bra}[1]{\langle #1 \vert}
\newcommand{\ket}[1]{\vert #1 \rangle}

\newcommand{\proj}[2]{\left\vert #1 \right\rangle \!\left\langle #2 \right\vert }
\newcommand{\iproj}[2]{\vert #1^{(i)} \rangle \langle #2^{(i)}\vert}
\newcommand{\expt}[1]{\left\langle #1 \right\rangle} %expectation value
\newcommand{\e}{\mathrm{e}}

\newcommand{\mr}{\mathrm}
\newcommand{\mc}{\mathcal}
\newcommand{\hc}{\mathrm{H.c.}}
\let\dag\dagger
\let\w\omega
\let\W\Omega
\let\mr\mathrm
\newcommand{\geff}{g_{\mr{eff}}}
\newcommand{\dket}[1]{\vert #1 \rangle\!\rangle}
\newcommand{\dbra}[1]{\langle\!\langle #1 \vert}
\newcommand{\Rb}{{}^{87}\mathrm{Rb}}

\begin{document}
\title{M{\o}lmer-S{\o}rensen entangling gate for cavity QED systems}
\author{Hiroki Takahashi}
\email[]{ht74@sussex.ac.uk}
\author{Pedro Nevado Serrano}
\author{Matthias Keller}
%\date{2015}
\affiliation{Department of Physics and Astronomy, University of Sussex,
Brighton, BN1 9QH, United Kingdom}

\begin{abstract} 
The M{\o}lmer-S{\o}rensen gate is a state-of-the-art entangling gate in the ion trap quantum computing where the gate fidelity can exceed 99\%. Here we propose an analogous implementation in the setting of cavity QED. The cavity photon mode acts as the bosonic degree of freedom in the gate in contrast of that played by a phonon mode in ion traps. This is made possible by utilising cavity assisted Raman transitions interconnecting the logical qubit states embedded in a four-level energy structure, making the ``anti-Jaynes-Cummings'' (AJC) term available under the rotating-wave approximation. We identify practical sources of infidelity and discuss their effects on the gate performance. Our proposal not only demonstrates an alternative entangling gate scheme but also sheds new light on the relationship between ion traps and cavity QED, in the sense that many techniques developed in the former are transferable to the latter through our framework.  
% This scheme has several favourable features in comparison with the ion trap implementation, such as no Lamb-Dicke parameter, no off-resonant couplings and no heating effects. 
% Applicability of the scheme is  not limited to single atoms or ions in a cavity but solid-state systems.  
\end{abstract}

\maketitle
 
\section{Introduction}
\label{sec:introduction}

Currently trapped atomic ions are among the most successful platforms for quantum information processing (QIP). A number of quantum algorithms \cite{Gulde2003,Brickman2005,monz2016realization}, entanglement of up to 14 ions \cite{Monz2011} and quantum simulation of spin systems \cite{Blatt2012,Islam2013} have been demonstrated, to name a few. Many of those achievements rely on the realization of high fidelity two-qubit entangling gates known as M{\o}lmer-S{\o}rensen (MS) \cite{Molmer1999} or the geometric phase gate \cite{Leibfried2003a} \footnote{The distinction between the M{\o}lmer-S{\o}rensen (MS) and geometric phase gate is usually made by the basis states they operate on. The MS-gate is in the form of $\sigma_x\otimes\sigma_x$ whereas the geometric phase gate is in $\sigma_z\otimes\sigma_z$. In particular the latter does not flip the logical qubit states. Following this convention we call our gate M{\o}lmer-S{\o}rensen due to the derived form of the Hamiltonian (\ref{eq:H_MS})}. These gates exploit a collective phonon mode shared by the ions to mediate a state-dependent force  and induce a quantum phase conditioned on the collective atomic states. Notable characteristics of this gate scheme are 1) individual addressing of the ions is not required, 2) the time evolution is cyclic such that the electronic and phonon degrees of freedom become disentangled at certain times and 3) the scheme is insensitive to the ions' initial motional state \cite{Sorensen2000}. Due to these favorable features, the gate can achieve a fidelity in excess of 99\% \cite{Benhelm2008a,ballance2016high}.

On the other hand, cavity QED is a paradigm where stationary quantum emitters (e.g. single atoms) interact with quantized radiation fields. It serves as a versatile platform for studies in quantum optics and for quantum information. Namely cavity QED systems are regarded as a vital building block in the development of quantum networks \cite{Kimble2008}. In the quantum network architecture, each network node is required to be a quantum register capable of multi-qubit quantum logic operations. Even though the recent experimental progress makes it possible to couple multiple qubits to a single optical cavity \cite{Casabone2015,Reimann2015,neuzner2016interference,begley2016optimized}, entangling gate operations within a single cavity QED system have not been demonstrated despite a number of  theoretical proposals \cite{Pellizzari1995,Zheng2000,You2003,Zheng2004}.

In this article we propose an alternative implementation of a quantum entangling gate for cavity QED systems, which has a direct correspondence to the MS gate in ion traps. It is well known that trapped ions and cavity QED systems share similar physical compositions, i.e. effective spins coupled with a quantized bosonic mode \cite{Blockley1992}. The prime difference is that in ion traps both Jaynes-Cummings (JC) and anti-Jaynes-Cummings (AJC) Hamiltonians are naturally available by addressing the red and blue-sideband transitions of the ions respectively \cite{Leibfried2003}, whereas in cavity QED normally only the JC Hamiltonian is available. However this restriction can be lifted by utilizing two cavity-assisted Raman transitions interconnecting qubit states embedded in a four-level energy structure. This was first discovered in conjunction with the realization of the Dicke model \cite{Dimer2007,baden2014realization} and later used for studies of the Rabi model \cite{grimsmo2013cavity}, but it has never been discussed in terms of a quantum logic gate as per our knowledge. 

Even though our proposal is directly inspired by the MS-gate, there are notable differences from the ion-trap implementation. Firstly cavity QED systems are essentially a single-mode system as opposed to the inherent multiple mechanical modes in a string of ions. Therefore our scheme is free from the issues in ion traps such as off-resonant excitations to irrelevant mechanical modes and spectral congestion in the mode structure with the increasing number of ions. Along the same line, there is no Lamb-Dicke parameter in our scheme, which means there is no compromise between the spatial localization of the qubits and the gate speed. Finally, the optical mode of a cavity can be regarded as being at zero temperature without the need of additional cooling. Hence our scheme does not suffer from the heating of the bosonic mode as it is often problematic in ion trap QIPs. However, optical cavities normally have non-negligible field decay rates.

In the following we refer to the individual stationary qubits in the cavity as ``atoms'' for the sake of convenience. However, in an actual implementation they do not need to be single atomic particles. Indeed they could be e.g. molecules, nitrogen-vacancy centers in diamond or artificial atoms such as semiconductor quantum dots, as long as they have the required energy structure and transitions addressable by a cavity and external laser fields (see \ref{fig:level-scheme}). Therefore we expect that our proposal is relevant to a broad class of physical systems where direct interaction between the qubits is difficult to attain, but they can be indirectly coupled to each other via an optical cavity field.

The article is structured as follows: in Section \ref{sec:deriv-hamilt} we introduce the MS Hamiltonian. Section \ref{sec:effect-ac-stark} is devoted to the discussion of the cavity-induced Stark shift. The influence of the decay channels on the gate performance are presented in Section \ref{sec:cavity-atomic-decay}. Finally, in Section \ref{sec:cavity-atomic-decay} we summarize our main conclusions.

\section{Derivation of the Hamiltonian}
\label{sec:deriv-hamilt}

\begin{figure}[tb]
\centering
  \includegraphics[width=0.9\linewidth]{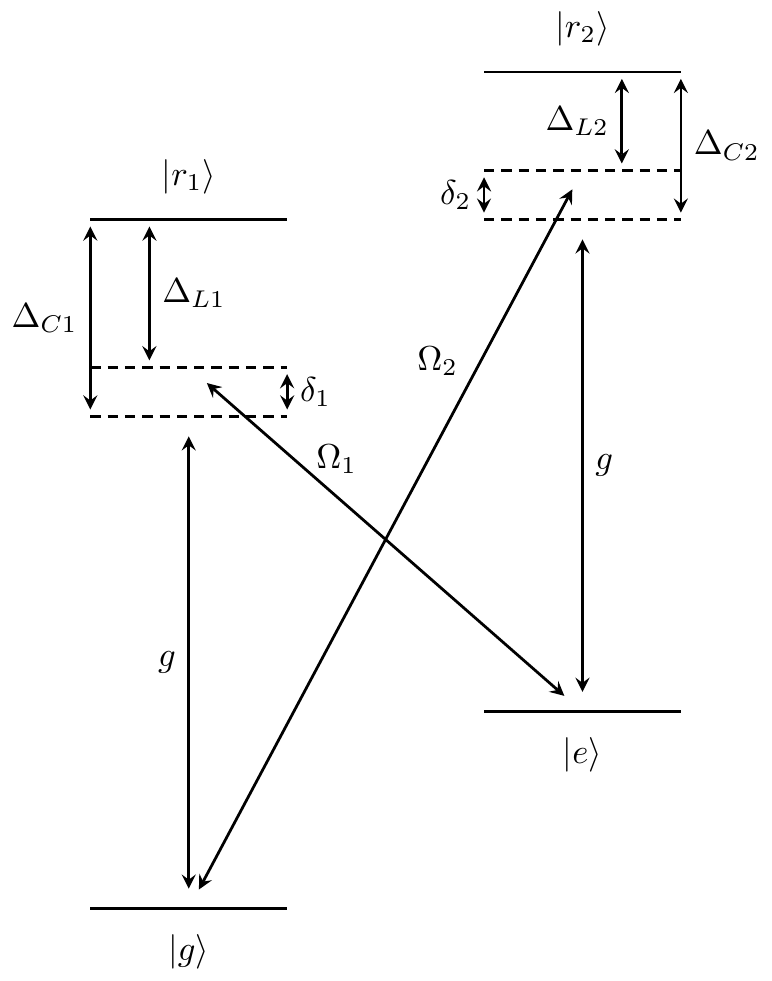}\caption{The level scheme of the atoms and transitions induced by the laser and cavity fields.}
\label{fig:level-scheme}
\end{figure}

We consider an ensemble of $N$ atoms coupled to a single cavity mode. All the atoms possess an identical four-level energy structure as shown in~\ref{fig:level-scheme}. The ground states $\ket{g}$ and $\ket{e}$ form a qubit whereas the excited states $\ket{r_1}$ and $\ket{r_2}$ mediate the coupling between the qubit states via cavity-assisted Raman transitions.  The cavity frequency $\w_c$ is near resonant with  the transitions $\ket{g} \leftrightarrow \ket{r_1}$ and $\ket{e} \leftrightarrow \ket{r_2}$ with corresponding detunings $\Delta_{C_1}$ and $\Delta_{C_2}$. We assume that the coupling to the cavity mode is uniform among the atoms and at the same strength of a vacuum Rabi frequency $2g$ for both $\ket{g} \leftrightarrow \ket{r_1}$ and $\ket{e} \leftrightarrow \ket{r_2}$. (The latter condition is not essential and can be relaxed.).
In addition, the atoms are externally driven by two laser fields. 
These two lasers off-resonantly drive the transitions $\ket{e} \leftrightarrow \ket{r_1}$ and $\ket{g} \leftrightarrow \ket{r_2}$ with Rabi frequencies $\Omega_1$ and $\Omega_2$, and detunings $\Delta_{L_1}$ and $\Delta_{L_2}$ respectively.
In this way a pair of Raman transitions is constructed to couple the qubit states. Each of them has a cavity induced transition on one arm  and a laser-induced transition on the other. The two-photon detunings $\delta_1$ and $\delta_2$ are given by 
\begin{align}
 \delta_1 &= \Delta_{C_1}-\Delta_{L_1}, \\
 \delta_2 &= \Delta_{C_2}-\Delta_{L_2}.
\end{align}
The Hamiltonian for the total atom-cavity system is composed of the bare energy $H_0$ and the interaction Hamiltonian $H_I$, i.e. $H = H_0 + H_I$. 
$H_0$ is (assuming $\hbar = 1$)
\begin{align}
 H_0 &= \w_{c} a^\dag a + \sum_{i=1}^{N}\left(\w_g\iproj{g}{g} + \w_e\iproj{e}{e}\right. \nonumber \\
 &\left.\quad+\w_{r_1}\iproj{r_1}{r_1}+\w_{r_2}\iproj{r_2}{r_2}\right).
\end{align}
Here, $a$ is the annihilation operator of the cavity photon, $\w_{\xi}$ and $\iproj{\xi}{\xi} \,(\xi=g,e,r_1,r_2)$ are the energy of the atomic level and the projector to the corresponding eigenstate of the $i$th atom respectively. 
On the other hand, $H_I$ is given by a sum of the interaction Hamiltonians for the individual atoms:
\begin{align}
 H_I &= \sum_{i=1}^{N} \left[\frac{\Omega_1}{2}\left(\e^{-i\w_1t}\iproj{r_1}{e}+\hc\right)\nonumber \right.\\
 &\left.+\frac{\Omega_2}{2}\left(\e^{-i\w_2t}\iproj{r_2}{g}+\hc\right) \right.\nonumber \\
 &\left.+g\left(a\iproj{r_1}{g}+a\iproj{r_2}{e}+\hc\right)\right].
\end{align}
Here, $\w_1$ and $\w_2$ are the optical frequencies of the driving lasers and $\hc$ indicates Hermitian conjugate of the preceding term inside the same bracket.
In the interaction picture with respect to the operator
\begin{align}
  H_1 &= \w_{c} a^\dag a+\sum_{i=1}^{N}\left(\w_g\iproj{g}{g} + \w_e\iproj{e}{e}\right. \nonumber \\
 &\left.\quad+\left(\w_{r_1}+\Delta_1\right)\iproj{r_1}{r_1}+\left(\w_{r_1}+\Delta_2\right)\iproj{r_2}{r_2}\right),\label{eq:H1}
\end{align}
with
\begin{align}
 \Delta_1 &= \frac{\Delta_{C_1}+\Delta_{L_1}}{2}, \\
 \Delta_2 &= \frac{\Delta_{C_2}+\Delta_{L_2}}{2},
\end{align}
%
%Then we get a new Hamiltonian $H'=U_1HU_1^{\dag}+iU_1^{\dag}\frac{dU_1}{dt}$ with $U_1 = \e^{iH_1t}$:
the new Hamiltonian, $H'=U_1HU_1^{\dag}+iU_1^{\dag}\frac{dU_1}{dt}$ with $U_1 = \e^{iH_1t}$, becomes
\begin{align}
 H' &= \sum_{i=1}^{N}\left[-\Delta_1\iproj{r_1}{r_1}-\Delta_2\iproj{r_2}{r_2}\right.\nonumber \\
 &+\frac{\Omega_1}{2}\left(\e^{i\frac{\delta_1}{2}t}\iproj{r_1}{e}+\hc\right) \nonumber \\
 &+\frac{\Omega_2}{2}\left(\e^{i\frac{\delta_2}{2}t}\iproj{r_2}{g}+\hc\right) \nonumber \\
 &\left.+g\left(\e^{-i\frac{\delta_1}{2}t}a\iproj{r_1}{g}+\e^{-i\frac{\delta_2}{2}t}a\iproj{r_1}{g}+\hc\right)\right]. \label{eq:H-prime}
\end{align}
Assuming
\begin{align}
 \abs{\Delta_{1,2}} \gg \abs{g}, \,\abs{\W_{1,2}},\,\abs{\delta_{1,2}},
 \label{eq:adiabatic-elimination}
\end{align}
the excited states $\ket{r_1}, \ket{r_2}$ can be adiabatically eliminated and we obtain an effective Hamiltonian:
\begin{align}
 H_{\mr{eff}} &= \frac{g^2}{2}\left(\frac{1}{\Delta_1}+\frac{1}{\Delta_2}\right)a^\dag a \nonumber\\
 &+\sum_{i=1}^{N} \left\{ \frac{\W_2^2}{4\Delta_2}\iproj{g}{g}+\frac{\W_1^2}{4\Delta_1}\iproj{e}{e}\right.\nonumber \\
 &+\frac{g^2}{2}\left(\frac{1}{\Delta_2}-\frac{1}{\Delta_1}\right)a^\dag a\left(\iproj{e}{e}-\iproj{g}{g}\right) \nonumber \\
 &\left.+\frac{g\W_1}{2\Delta_1}\left(\e^{-i\delta_1 t}a\iproj{e}{g}+\hc\right)\right.\nonumber \\
 &+\left. \frac{g\W_2}{2\Delta_2}\left(\e^{-i\delta_2 t}a\iproj{g}{e}+\hc\right)\right\}. \label{eq:Heff1}
\end{align}
The first two terms inside the sum over the atom index $i$ correspond to the ac Stark shifts caused by the driving lasers. These terms in addition to the first term in (\ref{eq:Heff1}) shifts the bare energy eigenfrequencies with constant offsets. Thus they can be removed from the equation by moving to another interaction picture with respect to
\begin{align}
 H_2 &= \frac{g^2}{2}\left(\frac{1}{\Delta_1}+\frac{1}{\Delta_2}\right)a^\dag a \nonumber\\
 &\quad+\sum_{i=1}^{N}\{ \frac{\W_2^2}{4\Delta_2}\iproj{g}{g}+\frac{\W_1^2}{4\Delta_1}\iproj{e}{e}\}
 \nonumber\\
 &=\Delta_c^{(s)}a^\dag a+\sum_i\{\Delta_g^{(s)}\iproj{g}{g}+\Delta_e^{(s)}\iproj{e}{e}\}, \label{eq:H2}
\end{align}
with
\begin{align}
 \Delta_c^{(s)} &\equiv \frac{g^2}{2}\left(\frac{1}{\Delta_1}+\frac{1}{\Delta_2}\right), \\
 \Delta_g^{(s)} &\equiv \frac{\W_2^2}{4\Delta_2}, \\
 \Delta_e^{(s)} &\equiv \frac{\W_1^2}{4\Delta_1},
\end{align}
resulting in
\begin{align}
 H'_{\mr{eff}} &= U_2H_{\mr{eff}}U_2^\dag - H_2 \nonumber \\
 &= \sum_{i=1}^{N} \left[\frac{g^2}{2}\left(\frac{1}{\Delta_2}-\frac{1}{\Delta_1}\right)a^\dag a\left(\iproj{e}{e}-\iproj{g}{g}\right)\right.\nonumber\\
 &\left.+\frac{g\W_1}{2\Delta_1}\left(\e^{-i\delta'_1 t}a\iproj{e}{g}+\hc\right)\right.\nonumber \\
 &+\left. \frac{g\W_2}{2\Delta_2}\left(\e^{-i\delta'_2 t}a\iproj{g}{e}+\hc\right)\right], \label{eq:Heff-prime}
\end{align}
where
\begin{align}
 \delta'_1 &\equiv \delta_1+\Delta_c^{(s)}+\Delta_g^{(s)}-\Delta_e^{(s)}, \\
 \delta'_2 &\equiv \delta_2+\Delta_c^{(s)}+\Delta_e^{(s)}-\Delta_g^{(s)}.
\end{align}
By setting
\begin{align}
 \delta'_1 = &\delta'_2 \equiv \delta, \label{eq:det-cond} \\
 \frac{\W_1}{\W_2} &= \frac{\Delta_1}{\Delta_2}, \label{eq:omega-cond} 
\end{align}
(\ref{eq:Heff-prime}) becomes
\begin{align}
 H'_{\mr{eff}} = \chi a^\dag a S_z + g_{\mr{eff}}(\e^{-i\delta t}a+\e^{i\delta t}a^\dag) S_x. \label{eq:Heff-prime2}
\end{align}
Here we have defined the following constants and operators:
\begin{align}
  \chi &\equiv g^2\left(\frac{1}{\Delta_1}-\frac{1}{\Delta_2}\right), \label{eq:chi} \\
 \geff &\equiv \frac{g\W_1}{\Delta_1} = \frac{g\W_2}{\Delta_2}, \label{eq:geff} \\
 S_z &\equiv \frac{1}{2}\sum_{i=1}^{N} (\iproj{e}{e}-\iproj{g}{g}), \label{eq:Sz} \\
 S_x &\equiv \frac{1}{2} \sum_{i=1}^{N} (\iproj{e}{g}+\iproj{g}{e}). \label{eq:Sx}
\end{align}
The second term in (\ref{eq:Heff-prime2}) is the desired M{\o}lmer-S{\o}rensen interaction:
\begin{align}
 H_{\mr{MS}} = g_{\mr{eff}}(\e^{-i\delta t}a+\e^{i\delta t}a^\dag) S_x. \label{eq:H_MS}
\end{align}
It is known that the integral of this Hamiltonian, denoted here $U_{\mr{MS}}(t)$, can be exactly calculated \cite{Sorensen2000,Roos2008}.  
\begin{align}
 U_{\mr{MS}}(t) &= \e^{-i(\alpha(t)a^\dag+\alpha^\ast(t) a)S_x}\e^{i\beta(t)S_x^2},\label{eq:U-MS}
\end{align}
where
\begin{align}
 \alpha(t) &= i\frac{\geff}{\delta}(1-\e^{i\delta t}), \label{eq:alpha}\\
 \beta(t) &= \left(\frac{\geff}{\delta}\right)^2(\delta t-\sin\delta t) \label{eq:beta}.
\end{align}
\begin{figure}[t!]
 \centering
  \includegraphics[width=\linewidth]{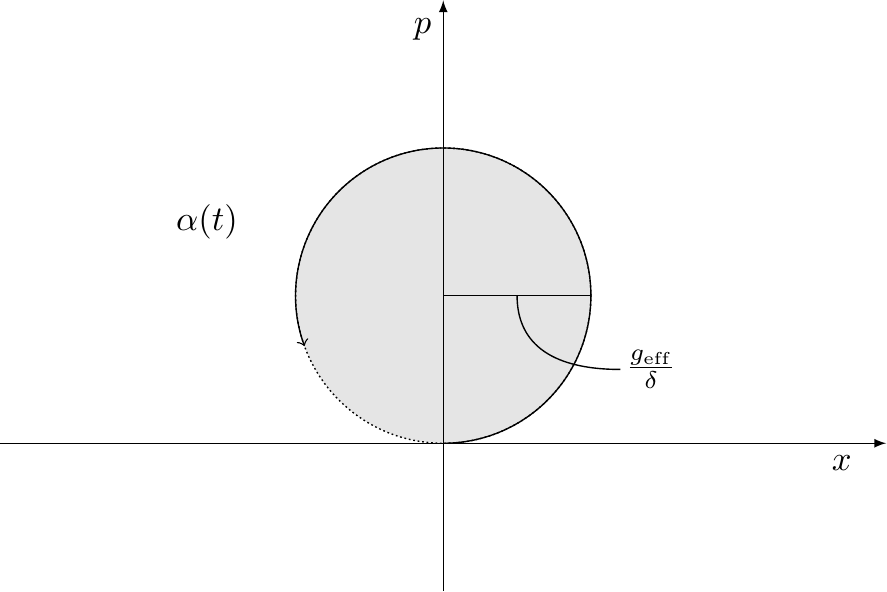}
 \caption{The trajectory of $\alpha(t)$ in phase space when $\geff, \delta > 0$.}
 \label{fig:alpha-t}
\end{figure}
$\alpha(t)$ draws a circle with a radius of $\frac{\geff}{\abs{\delta}}$ in phase space as seen in \ref{fig:alpha-t} and it returns to the origin after every $\tau = \frac{2\pi}{\abs{\delta}}$. Therefore at $t = n\tau\,(n=0, 1, 2,\dots )$, $\alpha(t) = 0$ and $U_{\mr{MS}}$ becomes a propagator involving only the atomic degrees of freedom:
\begin{align}
 U_{\mr{MS}}(t = n\tau) &= \e^{i\beta_nS_x^2}
\end{align}
where $\beta_n=2n\pi(\frac{\geff}{\delta})^2\mr{sign}(\delta)$ can be expressed by the area $A$ enclosed by $\alpha(t)$ as $\beta_n = 2n\,\mr{sign}(\delta)A$, and hence is called geometric phase. By choosing the two photon detuning such that
\begin{align}
 \abs{\delta} = 2\sqrt{m}\,\geff \quad (m = 1, 2, 3, \dots), \label{eq:delta-definition}
\end{align}
the geometric phase becomes $\beta_m = \mr{sign}(\delta)\frac{\pi}{2}$ at $t =m\tau$ and  
$U_{\mr{MS}}$ can be used to generate maximally entangled states \cite{Molmer1999}. We define the gate time
\begin{align}
 t_{\mr{gate}} = m\tau = \frac{\pi\abs{\delta}}{2\geff^2}.\label{eq:tgate}  
\end{align}
In particular when $N=2$, it accomplishes the following transformations of the two-qubit basis states, which are equivalent to the controlled-not gate up to single-qubit rotations: 
\begin{equation}
 \begin{aligned}
  \ket{\phi_1} \equiv \ket{gg} &\rightarrow \ket{\Phi_1^{(\pm)}} = \frac{1}{\sqrt{2}}(\ket{gg}\pm i\ket{ee}), \\
  \ket{\phi_2} \equiv \ket{ge} &\rightarrow \ket{\Phi_2^{(\pm)}} = \frac{1}{\sqrt{2}}(\ket{ge}\pm i\ket{eg}), \\
  \ket{\phi_3} \equiv \ket{eg} &\rightarrow \ket{\Phi_3^{(\pm)}} = \frac{1}{\sqrt{2}}(\pm i\ket{ge}+\ket{eg}), \\
  \ket{\phi_4} \equiv \ket{ee} &\rightarrow \ket{\Phi_4^{(\pm)}} = \frac{1}{\sqrt{2}}(\pm i\ket{gg}+\ket{ee}).
 \end{aligned}
 \label{eq:MS-gate-two-atoms}
\end{equation}
Here the plus and minus signs correspond to the sign of $\delta$.

So far we have neglected the first term in (\ref{eq:Heff-prime2}) which represents the differential ac Stark shift induced by the cavity field:
\begin{align}
 H_{\mr{AS}} = \chi a^\dag a S_z.
\end{align}
This Hamiltonian can cause a deviation from the ideal time evolution of $H_{\mr{MS}}$.
By setting $\Delta_1=\Delta_2$ in addition to the conditions (\ref{eq:det-cond}) and (\ref{eq:omega-cond}), which in turn means $\W_1 = \W_2$, $\chi$ vanishes. However in general this additional condition may not be satisfied since it imposes a constraint $\w_{r_1}-\w_{g} \approx \w_{r_2}-\w_{e}$ on the energy structure if $\abs{\delta} \ll \abs{\Delta_1}, \abs{\Delta_2}$. When this is not the case,  the caused deviation may not be negligible depending on the magnitude of $\chi$. 
Another possible deviation could arise from dissipative processes such as the cavity field decay and atomic spontaneous emissions from the excited states. %In contrast with ion traps, the optical fields in cavity QED systems suffer inevitable scattering or transmission losses at the cavity mirrors. 
% In order to incorporate these dissipations, we need to use a master equation for the time evolution:
% %
% \begin{align}
%  \frac{d\rho}{dt} &= -i[H'_{\mr{eff}}, \rho] + \kappa(2a\rho a^\dag-a^\dag a\rho- \rho a^\dag a)
% \end{align}
% %
% where $\rho$ is the density operator for the total system and $\kappa$ is the field dissipation rate.

In the following sections, we treat these two kinds of imperfections -- the effect of $\chi$ and that of dissipative processes -- in the case of two atoms.

\section{Effect of the cavity-induced ac Stark shift}
\label{sec:effect-ac-stark}

\begin{figure}[t!]
 \centering
  \includegraphics[width=0.8\linewidth]{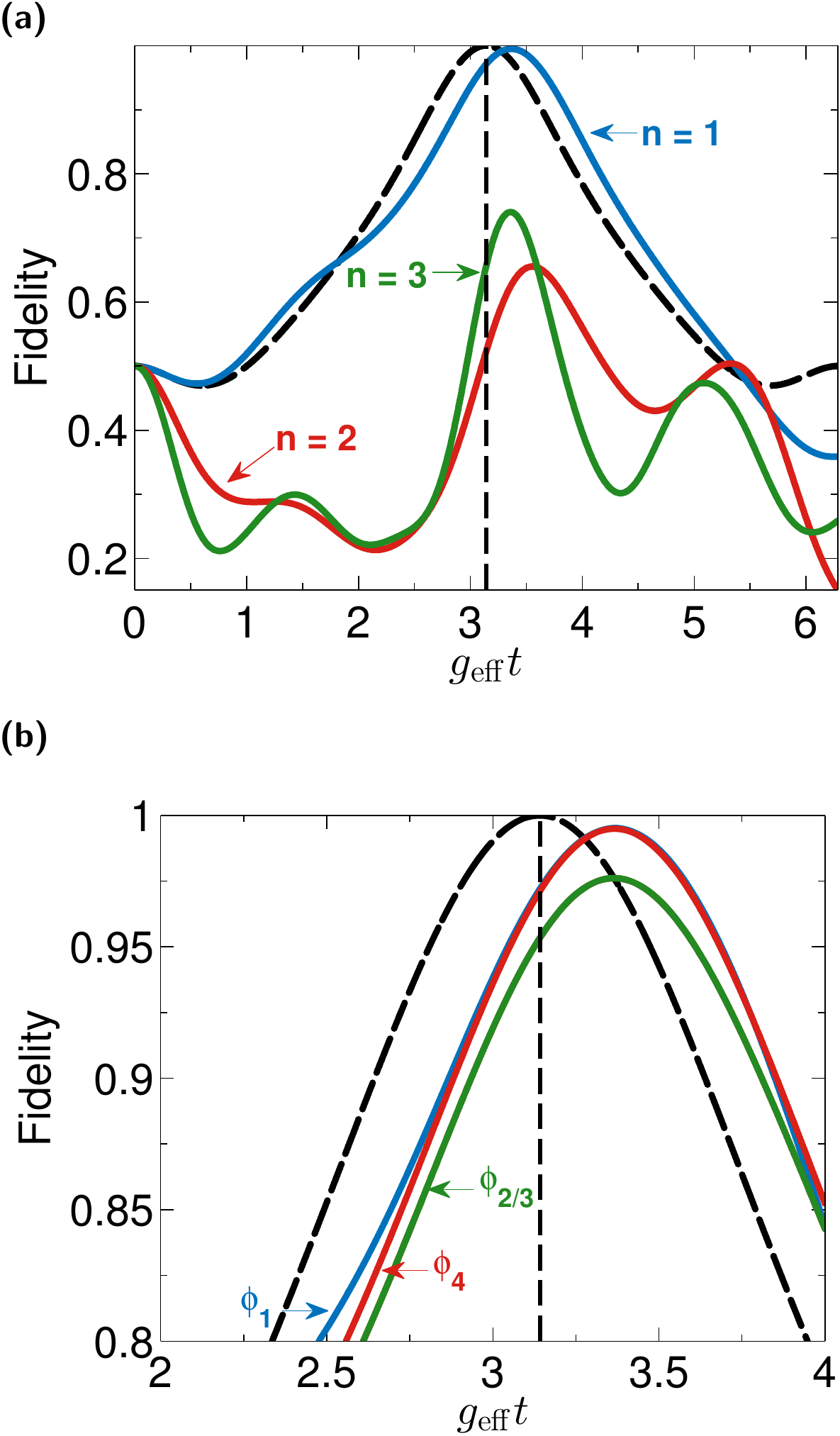}
  \caption{\textbf{(a)} Fidelity to $\ket{\Phi_1^{(+)}}$ for the initial state $\ket{\phi_1}\ket{n}$ as a function of time. $\chi = 0$  for the dashed black curve and $\chi/\geff = 0.5$ for the others. $\delta/\geff = 2$ for all the plots. The initial cavity photon numbers are $n = 0$ (blue), 1 (red) and 2 (green) respectively. The vertical dashed line shows the time at which the maximum fidelity is attained in the $\chi = 0 $ case. \textbf{(b)} Fidelities to the corresponding $\ket{\Phi_i^{(+)}}$ for different atomic initial states with the cavity state prepared in the vacuum. The solid curves correspond to the initial states of $\ket{\phi_1}\ket{0}$ (blue), $\ket{\phi_4}\ket{0}$ (red) and $\ket{\phi_2}\ket{0}$ or $\ket{\phi_3}\ket{0}$ (green) respectively. The black dashed curve is the fidelity with $\chi = 0$ which depends on neither initial atomic or cavity states.}
 \label{fig:initial-states}
\end{figure}

When $\chi \neq 0$, inclusion of $H_{\mr{AS}}$ results in deviations from the ideal entangling gate operations (\ref{eq:MS-gate-two-atoms}).
Let us denote the propagator of Hamiltonian (\ref{eq:Heff-prime2}) by $V(t)$. Then the wave function at time $t$ is given by
\begin{align}
 \ket{\Psi(t)} = V(t)\ket{\Psi(0)}, \label{eq:time-evolution-V}
\end{align}
where we assume that the initial state $\ket{\Psi(0)} = \ket{\phi_i}\ket{n}$ $(i = 1, 2, 3, 4)$, that is a tensor product of one of the logical basis states of two qubits (see (\ref{eq:MS-gate-two-atoms})) and a photon number state $\ket{n}$ with an arbitrary $n$. The fidelity of the state $\ket{\Psi(t)}$ with respect to the ideal gate output is 
\begin{align}
 F_{i,n}(t)& = \bra{\Phi_i^{(\pm)}}\mr{tr}_{\mr{photon}}(\proj{\Psi(t)}{\Psi(t)})\ket{\Phi_i^{(\pm)}} \nonumber\\
 = &\bra{\Phi_i^{(\pm)}}\mr{tr}_{\mr{photon}}(V(t)\proj{\phi_i}{\phi_i}\otimes\proj{n}{n}V^\dag(t))\ket{\Phi_i^{(\pm)}}. \label{eq:state-fidelity}
\end{align}
Here $\mr{tr}_{\mr{photon}}$ is a partial trace over the photon degree of freedom and $\ket{\Phi_i^{(\pm)}}$ is one of the atomic states for the ideal gate operation shown in (\ref{eq:MS-gate-two-atoms}), chosen accordingly to the initial state $\ket{\phi_i}$. The subscripts $i$ and $n$ represents the initial atomic state and photon number respectively.

Since $H_{\mr{AS}}$ is proportional to the photon number operator $a^\dagger a$, the perturbation caused by this Hamiltonian is expected to increase with the number of photons in the cavity. \ref{fig:initial-states}a shows time evolutions of $F_{1,n}(t)$, i.e. state fidelity to $\ket{\Phi_1^{(+)}}$ when the initial state is prepared in $\ket{\phi_1}\ket{n}$. Here we set $\delta = 2\geff > 0$.
When $\chi = 0$, the fidelity reaches the maximum value of unity at $\geff t = \pi$ (black dashed curve) and this behavior does not depend on the initial cavity state. However when $\chi \neq 0$, the fidelity shows strong dependence on the number of photons in the initial cavity state.
In particular, we see that there is an acute fidelity drop for cavity initial states $\ket{n}$ with $n > 0$. By preparing the initial cavity state in the vacuum state, the maximum fidelity remains close to unity even though the effect through dynamically generated photons is still present as a small drop of the fidelity and a time shift of the peak. Therefore in order to obtain the best gate performance for a given $\chi$, the cavity state has to be prepared close to the vacuum state.
%This spoils one of the virtues of the MS-gate that the gate performance does not depend on how the bosonic mode is prepared. Nonetheless it actually causes little problem for cavity QED systems in the optical domain.
However, since thermal excitations are negligible in the optical domain at the room temperature ($\expt{n} \sim 10^{-20}$), the cavity state remains essentially in vacuum unless we intentionally drive the cavity field. Hence this restriction to the initial cavity state causes little problem for cavity QED systems in the optical domain. This is in a stark contrast with ion traps where active cooling of phonon modes is always required.

In addition to the dependence on the initial cavity states, $H_{\mr{AS}}$ also leads to a dependence of the gate performance on the initial atomic states. \ref{fig:initial-states}b shows the fidelities for different atomic initial states to their target states. As can be seen, the initial states $\ket{\phi_2}$ and $\ket{\phi_3}$ are more sensitive to the perturbation than $\ket{\phi_1}$ and $\ket{\phi_4}$ when the cavity mode is prepared in vacuum.

These dependences of the gate performance on the initial photonic and atomic states when $\chi \neq 0$ can be illustrated by explicitly considering an approximation of the state fidelity for small $\chi$. In doing so, the main difficulty arises from the fact that $H_{\mr{MS}}$ and $H_{\mr{AS}}$ do not commute with each other. In \cite{grimsmo2013cavity}, the authors presented steady-state analysis of the same Hamiltonian (\ref{eq:Heff-prime2}) (or equivalently (\ref{eq:HI})). However, here we are interested in explicit time evolutions of states under the Hamiltonian. In Appendix~\ref{sec:pert-expans-prop} we show that in a certain interaction picture (defined by the relation (\ref{eq:Psi-PsiII})) the propagator can be perturbatively expanded in terms of a  Hamiltonian $H_{\mr{II}}(t)$ as shown in (\ref{eq:V-Dyson-series}). For the initial state $\ket{\Psi(0)} = \ket{\phi_i}\ket{n}$, we consider the time evolution of the state overlap with its target state $\ket{\Phi_i}\ket{n}$ (In the following, we assume $\delta > 0$ without loss of generality and omit $(+)$ in $\Phi_i^{(+)}$). That is
\begin{align}
 \eta_{i,n}(t) &= \bra{\Phi_i}\bra{n}U_{\mr{I}}^\dag(t)U_{\mr{II}}^\dag(t)V_{\mr{II}}(t)\ket{\phi_i}\ket{n}. \label{eq:state-overlap}
\end{align}
The corresponding state fidelity is given by $\abs{\eta_{i,n}(t)}^2$. Note that this fidelity is not exactly same as $F_{i,n}(t)$ given in (\ref{eq:state-fidelity}) as we did not trace out the photonic degree of freedom in (\ref{eq:state-overlap}). However as long as the additional excitation of photons due to $H_{\mr{AS}}$ \footnote{Even though $H_{\mr{AS}} \propto a^\dag a$ alone preserves the number of the cavity photons, due to the fact that $H_{\mr{AS}}$ and $H_{\mr{MS}}$ do not commute, the commutators between them arising in the propagator $V(t)$ causes a change of the cavity photons in addition to that caused by the MS-gate process.} is small for small $\chi$, $\abs{\eta_{i,n}(t)}^2$ is a good approximation of $F_{i,n}(t)$. According  to (\ref{eq:V-Dyson-series}), the state overlap $\eta_{i,n}(t)$ can be expanded as follows:
\begin{align}
 \eta_{i,n}(t) &= \eta_{i,n}^{(0)}(t) + \eta_{i,n}^{(1)}(t) + \eta_{i,n}^{(2)}(t) + \hdots, \label{eq:eta_expansion}
\end{align}
with
\begin{align}
 \eta_{i,n}^{(0)}(t)  &= \bra{\Phi_i}\bra{n}U_{\mr{I}}^\dag(t)U_{\mr{II}}^\dag(t)\ket{\phi_i}\ket{n}, \label{eq:eta0}\\
 \eta_{i,n}^{(1)}(t)  &= -i\bra{\Phi_i}\bra{n}U_{\mr{I}}^\dag(t)U_{\mr{II}}^\dag(t)\int_0^t\!H_\mr{II}(t')\,dt'\ket{\phi_i}\ket{n}, \label{eq:eta1}\\
 \eta_{i,n}^{(2)}(t) &\nonumber\\
 = (-i)^2&\bra{\Phi_i}\bra{n}U_{\mr{I}}^\dag(t)U_{\mr{II}}^\dag(t)\int_0^t\!\int_0^{t'}\!H_\mr{II}(t')H_\mr{II}(t'')\,dt'dt''\ket{\phi_i}\ket{n}. \label{eq:eta2}
\end{align}
$\eta_{i,n}^{(k)}$ is on the order of $\mc{O}(\chi^k)$ and $\eta_{i,n}^{(0)}(t)$ is the state overlap for the ideal MS-gate without the cavity induced ac-Stark shift. Using the energy eigenstates of $H_{MS}'$ given in (\ref{eq:eigenstate-Sx1})--(\ref{eq:eigenstate_S0-Sx0}) (see also (\ref{eq:def-jn-ket}) for the definition of the double-bracket state), the initial states $\ket{\phi_i}\ket{n}$ can be written as 
\begin{align}
\ket{\phi_{1,4}}\ket{n} &= \frac{1}{2}\sum_{m}d_{mn}(-\alpha)\dket{1, m} \nonumber\\
 &\mp \frac{1}{\sqrt{2}}\dket{0, n} %\nonumber\\
 + \frac{1}{2}\sum_{m}d_{mn}(\alpha)\dket{-1, m}, \label{eq:phi14-in-eigenstates}\\
\ket{\phi_{2,3}}\ket{n} &= \frac{1}{2}\sum_{m}d_{mn}(-\alpha)\dket{1, m} \nonumber\\
 \pm \frac{1}{\sqrt{2}}&\ket{S=0, S_x=0}\ket{n} %\nonumber\\
 - \frac{1}{2}\sum_{m}d_{mn}(\alpha)\dket{-1, m}, \label{eq:phi23-in-eigenstates}
\end{align}
where $d_{mn}(\pm\alpha) = \bra{m}D(\pm\alpha)\ket{n}$ and the plus and minus signs in front of the second term in (\ref{eq:phi14-in-eigenstates}) differentiate $\ket{\phi_4}$ and $\ket{\phi_1}$ respectively, and similarly $\ket{\phi_2}$ (plus) and $\ket{\phi_3}$ (minus) in (\ref{eq:phi23-in-eigenstates}). Likewise the target states $\ket{\Phi_i}\ket{n}$ are
\begin{align}
 \ket{\Phi_{1,4}}\ket{n} &=\frac{\e^{i\frac{\pi}{4}}}{2}\sum_{m}d_{mn}(-\alpha)\dket{1, m} \nonumber\\
 &\mp \frac{\e^{-i\frac{\pi}{4}}}{\sqrt{2}}\dket{0, n} %\nonumber\\
 + \frac{\e^{i\frac{\pi}{4}}}{2}\sum_{m}d_{mn}(\alpha)\dket{-1, m}, \label{eq:Phi14-in-eigenstates}\\
 \ket{\Phi_{2,3}}\ket{n} &= \frac{\e^{i\frac{\pi}{4}}}{2}\sum_{m}d_{mn}(-\alpha)\dket{1, m} \nonumber\\
 \pm \frac{\e^{-i\frac{\pi}{4}}}{\sqrt{2}}&\ket{S=0, S_x=0}\ket{n} %\nonumber\\
 - \frac{\e^{i\frac{\pi}{4}}}{2}\sum_{m}d_{mn}(\alpha)\dket{-1, m}. \label{eq:Phi23-in-eigenstates}
\end{align}
In terms of these energy eigenstates, the ideal MS-gate without the cavity ac-Stark shift can be understood solely by evolution of quantum phases in the interaction picture: during the ideal MS gate operation, each term in the form of $\dket{j, n}$ $(j = 0, \pm 1)$ in (\ref{eq:phi14-in-eigenstates}) and (\ref{eq:phi23-in-eigenstates}) acquire a phase $-\delta(n-(j\alpha)^2)t$. Note that $\ket{S=0, S_x=0}\ket{n}$ does not change because $j = n = 0$. At a time $t = t_{\mr{gate}}$, the part of this phase originating from photons ($=-n\delta t$) becomes an integer multiple of $2\pi$ whereas the spin-dependent phase ($= (j\alpha)^2\delta$) produces the relative phase required for $\ket{\Phi_{i}}\ket{n}$ (i = 1, 2, 3, 4) in  (\ref{eq:Phi14-in-eigenstates}) and (\ref{eq:Phi23-in-eigenstates}) if the condition (\ref{eq:delta-definition}) is satisfied. If the cavity ac-Stark shift is present, it disturbs this phase evolution by inducing transitions between $\dket{j, n}$ and $\dket{j', n'}$ where $\abs{j-j'} = 1$ (see \ref{eq:HII-expansion}).

Using expansions such as (\ref{eq:HII-expansion}) and (\ref{eq:HII2-expansion}), $\eta_{i,n}^{(k)}$ of any order $k$ can be explicitly calculated. It can be seen that $\eta_{i,n}^{(1)}$ is zero for $i = 2$ and $3$ for any $n$:
\begin{align}
 \eta_{2,n}^{(1)} = \eta_{3,n}^{(1)} = 0.
\end{align}
This is because both (\ref{eq:phi23-in-eigenstates}) and (\ref{eq:Phi23-in-eigenstates}) do not contain $\dket{0, m}$ for any $m$ whereas $H_{\mr{II}}$ only consists of off-diagonal terms in the form of $\dket{0,m}\dbra{\pm 1,n}$ and their conjugates (see (\ref{eq:HII-matirx1})~--~(\ref{eq:HII-matrix4})).
For $i = 1$ and $4$, we get
\begin{align}
 \eta_{1,n}^{(1)} &= \frac{n\chi\e^{i\delta n}}{2\sqrt{2}}\sum_m
 \frac{\e^{-iE_{1m}t}-\e^{-iE_{0n}t}}{E_{0n}-E_{1m}}\nonumber\\
 &\qquad\times(\abs{d_{mn}(\alpha)}^2+\abs{d_{mn}(-\alpha)}^2),\\
 \eta_{4,n}^{(1)} &= -\eta_{1,n}^{(1)}.
\end{align}
It is clear that both $\eta_{1,n}^{(1)}$ and $\eta_{4,n}^{(1)}$ become zero if $n =0$. Therefore if the cavity state is prepared in the vacuum ($n = 0$), the first order term in (\ref{eq:eta_expansion}) is zero irrespective of the initial atomic states. This argument can be extended to higher odd orders and one can find that $\eta_{i, 0}^{(k)} = 0$ is satisfied  for any $i$ if $k$ is an odd integer. If $n \ne 0$, $\eta_{i, n}^{(k)}$ is generally non-zero for $i = 1$ and $4$. These observations endorse the relative resilience of the fidelity to the perturbation caused by $H_{\mr{AS}}$ in the case of $n = 0$ as previously illustrated in \ref{fig:initial-states}a.
\begin{figure}[t!]
 \begin{center}
  \includegraphics[width=0.8\linewidth]{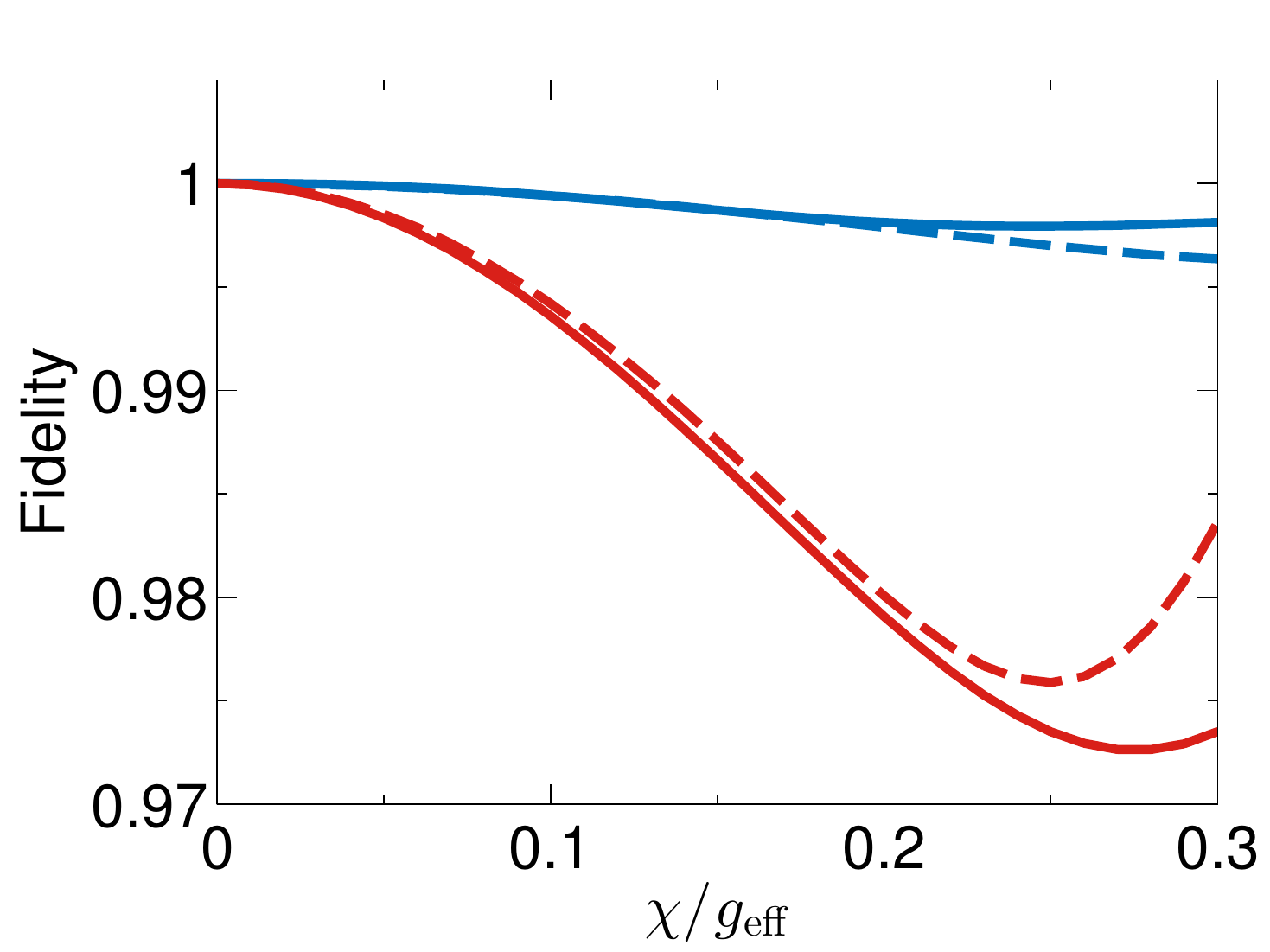}
  \caption{Maximally attainable state fidelities as a function of $\chi$. The solid lines are exact maximal fidelities based on (\ref{eq:state-fidelity}). The dashed lines are approximated fidelities based on (\ref{eq:eta_expansion}) up to the second order. The two colors indicate different initial states $\ket{\phi_1}\ket{0}$ (blue, upper two lines) and $\ket{\phi_2}\ket{0}$ (red, lower two lines). For all the lines, $\delta = 4g_{\mr{eff}}$.}
 \label{fig:state-fidelity-vs-chi}
 \end{center}
\end{figure}

Now we move on to the second order correction (\ref{eq:eta2}) for the case that the cavity is initially prepared in the vacuum state. 
By using (\ref{eq:phi14-in-eigenstates})--~(\ref{eq:Phi23-in-eigenstates}) and (\ref{eq:HII2-expansion}), we find
\begin{align}
 \eta_{1, 0}^{(2)} &= -\frac{\chi^2\e^{-i\frac{\pi}{4}}}{4}\sum_{l,m,n}l^2\e^{-iE_{1m}t}((-1)^{m+n}(1+(-1)^l))\nonumber\\
 &\times d_{m0}(\alpha)d_{n0}(\alpha)d_{lm}(\alpha)d_{ln}(\alpha)Y_{lmn}(t), \label{eq:eta2_10}\\
 \eta_{2, 0}^{(2)} &= -\frac{\chi^2\e^{-i\frac{\pi}{4}}}{4}\sum_{l,m,n}l^2\e^{-iE_{1m}t}((-1)^{m+n}(1-(-1)^l))\nonumber\\
 &\times d_{m0}(\alpha)d_{n0}(\alpha)d_{lm}(\alpha)d_{ln}(\alpha)Y_{lmn}(t), \label{eq:eta2_20}\\
 &\qquad\eta_{4, 0}^{(2)} = \eta_{1, 0}^{(2)}, \label{eq:eta2_40}\\
 &\qquad\eta_{3, 0}^{(2)} = \eta_{2, 0}^{(2)}, \label{eq:eta2_30}
\end{align}
where
\begin{align}
 Y_{lmn}(t) &= \int_0^t\!\int_0^{t'}\e^{-i(E_{0l}-E_{1m})t'}\e^{i(E_{0l}-E_{1m})t''} dt'dt'' \nonumber\\
 =\frac{1}{E_{1n}-E_{0l}}&\left(\frac{\e^{i(E_{1m}-E_{1n})t}-1}{E_{1m}-E_{1n}} + \frac{\e^{-i(E_{0l}-E_{1m})t}-1}{E_{0l}-E_{1m}} \right). \label{eq:Ylmn}
\end{align}

In order to assess the relative significance of $\eta_{i, 0}^{(2)}$ $(i = 1, 2, 3, 4)$ with respect to each other, we consider their functional dependences on $\alpha$.
Since $d_{nm}(\alpha) = O(\alpha^{\abs{n-m}})$, $d_{m0}(\alpha)d_{n0}(\alpha)d_{lm}(\alpha)d_{ln}(\alpha)$  is $O(\alpha^{m+n+\abs{m-l}+\abs{n-l}})$.
Hence, by considering possible combinations of $l$, $m$ and $n$ in (\ref{eq:eta2_10}) and (\ref{eq:eta2_20}), it can be found that at the lowest order in $\alpha$,
\begin{align}
 \eta_{1, 0}^{(2)} = \eta_{4, 0}^{(2)} &= O(\alpha^4), \label{eq:alpha_order_eta2_10+40}\\
 \eta_{2, 0}^{(2)} = \eta_{3, 0}^{(2)} &= O(\alpha^2). \label{eq:alpha_order_eta2_20+30}
\end{align}
The difference between (\ref{eq:alpha_order_eta2_10+40}) and (\ref{eq:alpha_order_eta2_20+30}) originates from the factors $(1\pm (-1)^l)$ in (\ref{eq:eta2_10}) and (\ref{eq:eta2_20}). Since $\abs{\alpha} = g_{\mr{eff}}/\delta < 1$ (see (\ref{eq:delta-definition})), this difference indicates that the second order correction is smaller for the initial states $\ket{\phi_{1,4}}\ket{0}$ than $\ket{\phi_{2,3}}\ket{0}$ , which is also consistent with \ref{fig:initial-states}b. 
Note that in \ref{fig:initial-states}b there is a small difference between the initial states $\ket{\phi_{1}}\ket{0}$ (blue) and $\ket{\phi_{4}}\ket{0}$ (red) whereas calculations based on (\ref{eq:eta2_10}) and (\ref{eq:eta2_40}) and even higher order terms in (\ref{eq:eta_expansion}) predict that they should exactly coincide. This is due to the partial trace of the photonic degree of freedom carried out in (\ref{eq:state-fidelity}) which is missing in the calculation of the state overlap (\ref{eq:state-overlap}).

\ref{fig:state-fidelity-vs-chi} shows attainable maximal state fidelities as a function of $\chi$ for different initial states. For the blue and red solid lines we numerically calculated $\mr{max}(F_{1,0}(t))$ and $\mr{max}(F_{2,0}(t))$ respectively by solving the Shr{\"{o}}dinger equations where the maximum is taken over time $t$ for a given $\chi$. On the other hand the dashed lines are approximated fidelities using (\ref{eq:eta_expansion}) up to the second order. That is, $\mr{max}\left(\abs{\eta^{(0)}_{1,0}(t)+\eta^{(2)}_{1,0}(t)}^2\right)$ for the blue dashed line and $\mr{max}\left(\abs{\eta^{(0)}_{2,0}(t)+\eta^{(2)}_{2,0}(t)}^2\right)$ for the red dashed line.
They reproduce the behaviors of the exact solutions, namely the distinction between different initial states, well up to $\chi \sim 0.2$ with an error in fidelity less than $10^{-3}$, confirming the validity of our perturbative approach.

\begin{figure}[t!]
 \begin{center}
  \includegraphics[width=0.8\linewidth]{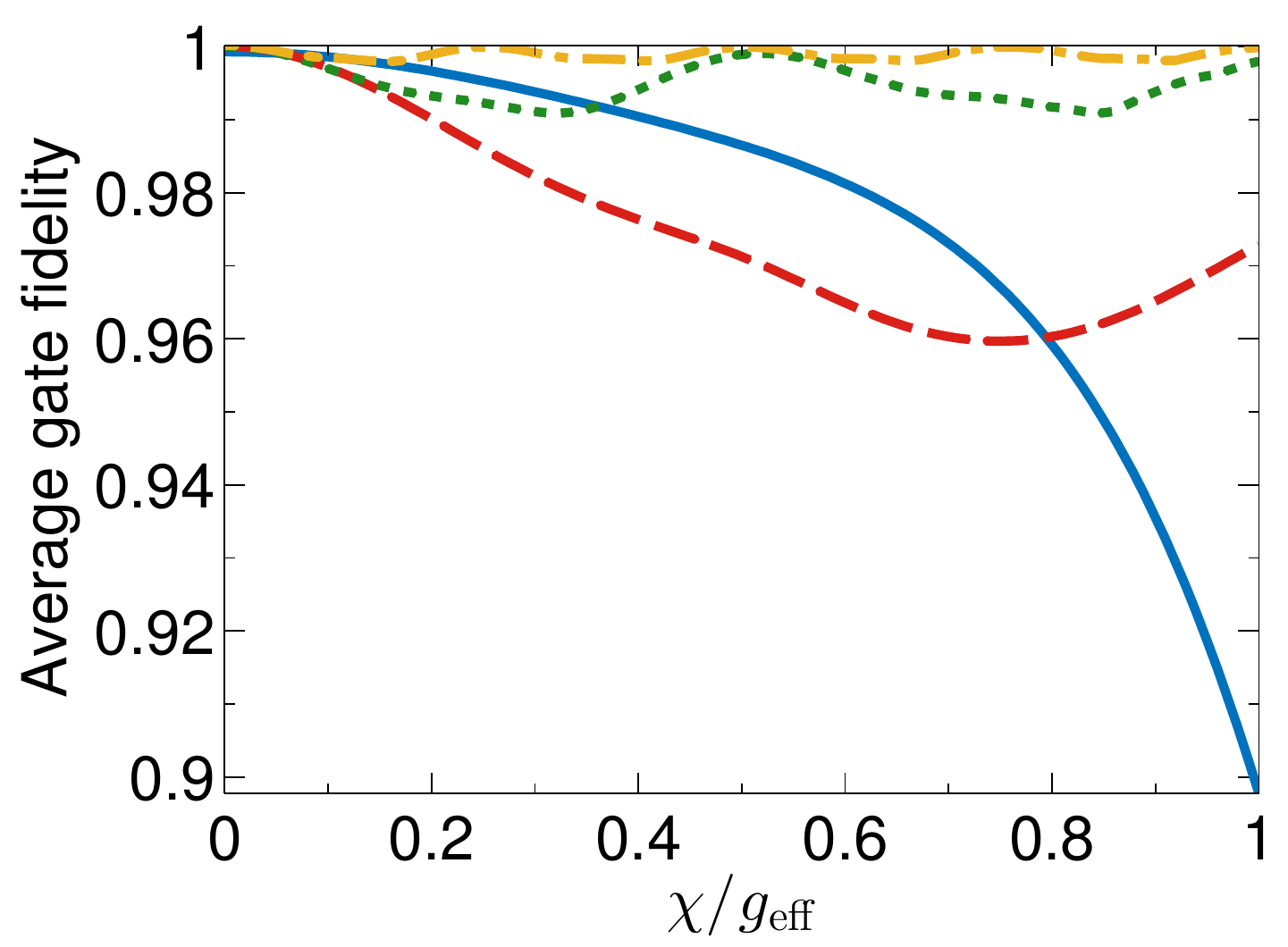}
  \caption{ Average gate fidelity as a function of $\chi$. $\delta/g_\mr{eff} = $ 2 (solid blue), 4 (dashed red), 8 (dotted green) and 16 (dash-dotted yellow).}
  \label{fig:avg_gate_fidelity}
\end{center}
\end{figure}
As the state fidelities to the target states differ among different initial sates in the presence of $H_{AS}$, the performance of the gate as a whole is quantified by the fidelity averaged over all possible initial states, called average gate fidelity. According to \cite{Nielsen2002}, the average gate fidelity is calculated using the following formula.
\begin{align}
 \bar{F}(t) = \frac{\sum_{i, j}\mr{tr}(U_\mr{MS}(t)(\sigma_i\otimes\sigma_j)U_\mr{MS}^\dag(t)\mc{E}(\sigma_i\otimes\sigma_j))+16}{80}, \label{eq:avg-gate-fidelity}
\end{align}
where $\sigma_{i,j}$ is one of the Pauli matrices or identity matrix and the sum runs through all possible combinations of those. $\mc{E}$ is a map between atomic density operators to represent the time evolution governed by $H_\mr{eff}'$:
\begin{align}
 \mc{E}(\rho) = \mr{tr}_{\mr{photon}}(V(t)(\rho\otimes\proj{0}{0})V^\dag(t)). 
\end{align}
Taking the maximum of (\ref{eq:avg-gate-fidelity}) in terms of time $t$, \ref{fig:avg_gate_fidelity} shows the attainable average gate fidelities as a function of $\chi$ for different two-photon detunings $\delta$. In the limit of $\delta/\geff \gg 1$, excitation of photons during the gate operation is increasingly suppressed as the radius of the trajectory (= $g_\mr{eff}/\delta$) in phase space reduces to zero. As a consequence, the perturbation by $H_\mr{AS}$ proportional to $a^\dag a$ becomes negligible such that the average gate fidelity remains close to unity with a large value of $\delta$ (see the dash-dotted yellow trace in \ref{fig:avg_gate_fidelity}). However this general trend does not necessarily apply to relatively small values of $\delta$ due to their oscillatory nature as seen in \ref{fig:avg_gate_fidelity}. For example the average gate fidelity for $\delta/\geff = 4$ is smaller than that for $\delta/\geff  = 2$ up to $\chi \sim 0.8$ .

\section{Cavity and atomic decay}
\label{sec:cavity-atomic-decay}

A deviation from the ideal gate operation could arise from dissipative processes such as the cavity decay and atomic spontaneous emissions. In this section, the effects of these processes are studied while assuming $\chi = 0$ for brevity.

The optical fields in cavity QED systems suffer inevitable scattering/absorption or transmission losses at the cavity mirrors. In order to incorporate these losses, we use a master equation for the time evolution:
\begin{align}
 \frac{d\rho}{dt} &= -i[H'_{\mr{eff}}, \rho] + \kappa(2a\rho a^\dag-a^\dag a\rho- \rho a^\dag a),
\end{align}
where $\rho$ is the density operator for the total system including the atomic and photonic degrees of freedom and $\kappa$ is the amplitude dissipation rate of the cavity field. Since the field dissipation does not depend on the atomic states, there is no difference between the average gate fidelity and the fidelity for a specific initial atomic state.
\begin{figure}[t!]
 \begin{center}
  \includegraphics[width=0.8\linewidth]{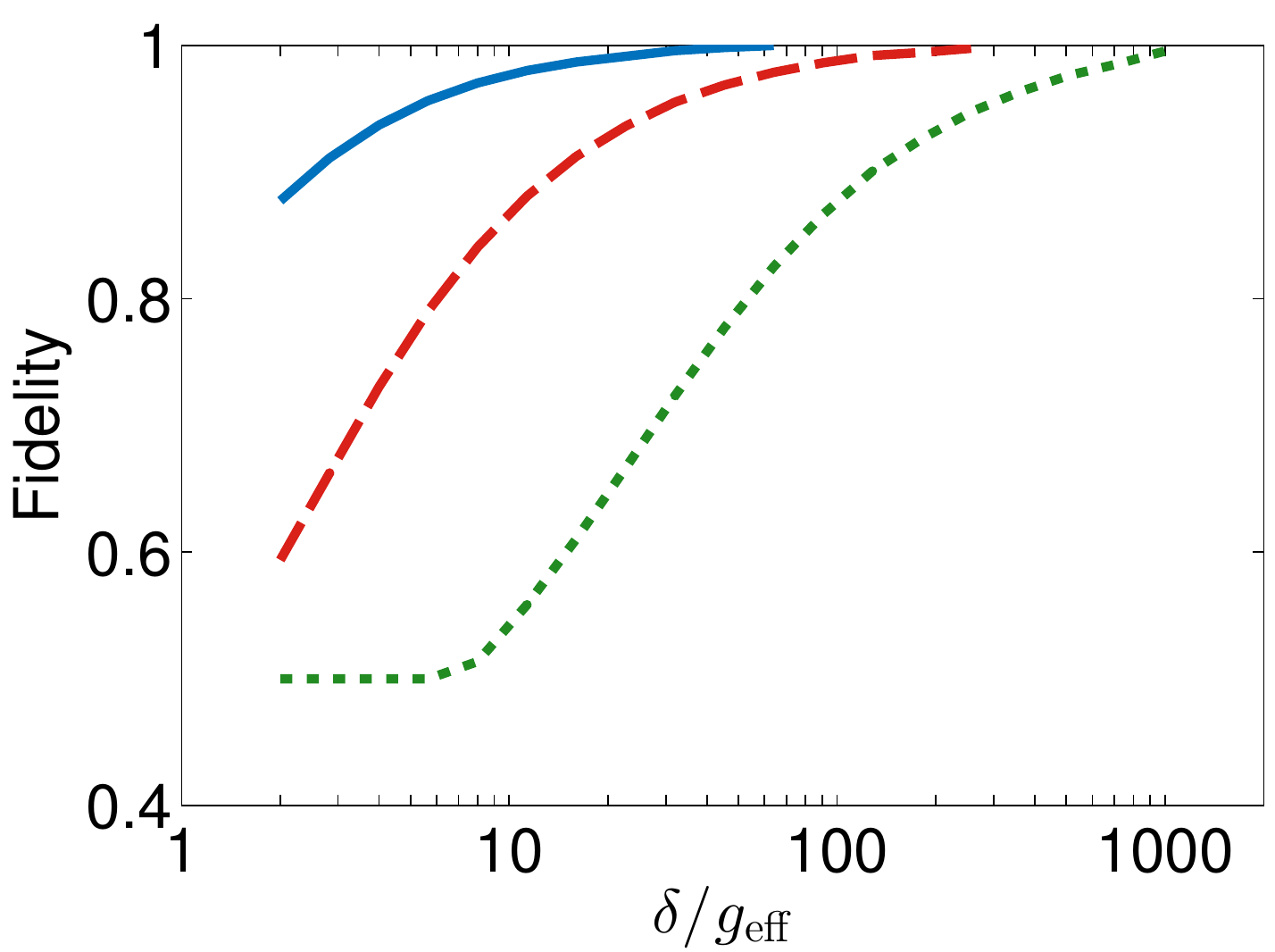}
  \caption{Gate fidelity as a function of $\delta$ when the cavity decay is present. $\kappa/\geff = $ 0.1 (solid blue), 1.0 (dashed red) and 10 (dotted green). }
  \label{fig:fidelity_kappa}
\end{center}
\end{figure}
\ref{fig:fidelity_kappa} shows attainable maximum fidelities for $\kappa \ne 0$. As can be seen, the detrimental effect of the field dissipation can be mitigated by increasing the two photon detuning $\delta$ with expense of increasing $t_\mr{gate}$. This can be understood as follows: in the regime where the photonic excitation is very small, the probability $P_{\kappa}$ that a photon loss occurs per unit time is proportional to the mean photon number $\sim O(\geff^2/\delta^2)$, which decreases quadratically with $\delta$. On the other hand $t_{\mr{gate}}$ only scales linearly with $\delta$ (see (\ref{eq:tgate})). Therefore the probability that a photon loss occurs within the gate time is $P_{\kappa}\,t_{\mr{gate}} \sim O(1/\delta)$ and hence increasing $\delta$ improves the achievable gate fidelity. 

The detunings required for a gate fidelity $\ge 0.99$ are $\delta/\geff \approx$ 20, 100 and 900 for $\kappa/\geff =$ 0.1, 1.0 and 10 respectively (see \ref{fig:fidelity_kappa}). Hence in principle the gate can be implemented with a high fidelity even in the regime where $\geff < \kappa$. However, since large detunings $\delta$ means long gate times $t_\mr{gate}$, increasing $\delta$ arbitrarily may not be an option if the system suffers from other decoherence mechanisms, such as the decay of atomic coherence, that cannot be mitigated by increasing $\delta$.
%Next we consider the effect of atomic spontaneous decay.

\begin{figure}[tb]
 \begin{center}
  \includegraphics[width=0.8\linewidth]{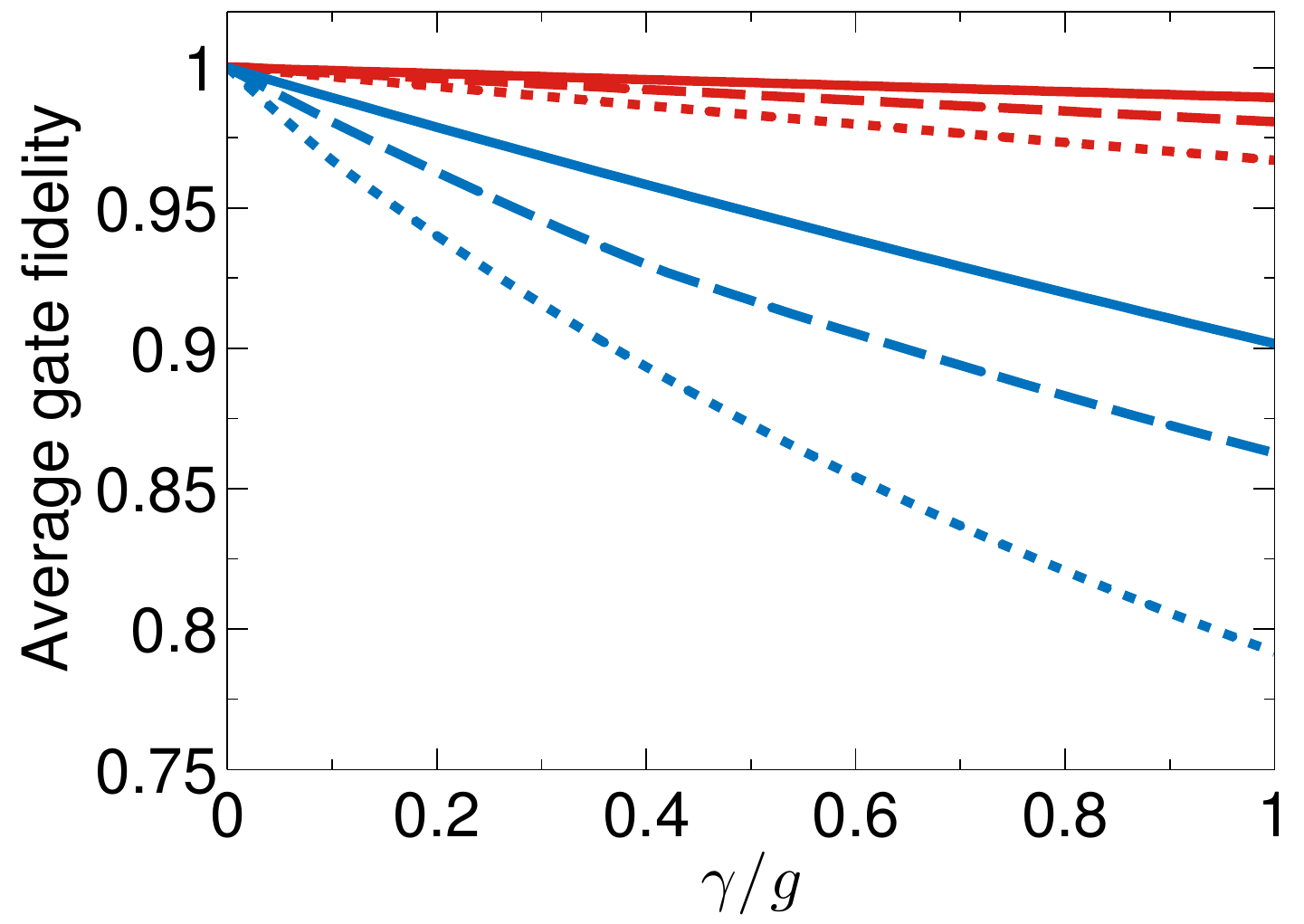}
  \caption{Average gate fidelity as a function of $\gamma$ normalized by $g$. Here we set $\Omega = g$. $\Delta/g = 100$ (red, upper three lines) and $10^3$ (blue, lower three lines). $\delta/\geff =$ 2 (solid), 50 (dashed) and 200 (dotted).}
  \label{fig:avg_fid_atomic_decay}
 \end{center}
\end{figure}

By adiabatically eliminating the excited states, the atomic spontaneous emissions from the excited states $\ket{r_1}$ and $\ket{r_2}$ can be effectively modeled with Lindblad operators acting on $\ket{e}$ and $\ket{g}$ as in the following master equation \cite{reiter2012effective}:
\begin{align}
 \frac{d\rho}{dt} &= -i[H'_{\mr{eff}}, \rho] + \sum_{i}\left(\mc{L}(C_{1g}^{(i)}, \rho) + \mc{L}(C_{1e}^{(i)}, \rho)\right. \nonumber \\
 &\quad\left. + \mc{L}(C_{2g}^{(i)}, \rho) + \mc{L}(C_{2e}^{(i)}, \rho)\right), \label{eq:master-eq-spont-decay}
\end{align}
where
\begin{align}
 \mc{L}(C, \rho) &\equiv 2\,C\rho C^\dag-C^\dag C\rho-\rho C^\dag C, \label{eq:Lindbladian}\\
 C_{1g}^{(i)} &\equiv \sqrt{\gamma_{1g}}(\frac{g}{\Delta_1}a\iproj{g}{g} + \frac{\Omega_1}{2\Delta_1}\iproj{g}{e}), \label{eq:def-C1g} \\
 C_{1e}^{(i)} &\equiv \sqrt{\gamma_{1e}}(\frac{g}{\Delta_1}a\iproj{e}{g} + \frac{\Omega_1}{2\Delta_1}\iproj{e}{e}), \label{eq:def-C1e} \\
 C_{2g}^{(i)} &\equiv \sqrt{\gamma_{2g}}(\frac{g}{\Delta_2}a\iproj{g}{e} + \frac{\Omega_2}{2\Delta_2}\iproj{g}{g}), \label{eq:def-C2g} \\
 C_{2e}^{(i)} &\equiv \sqrt{\gamma_{2e}}(\frac{g}{\Delta_2}a\iproj{e}{e} + \frac{\Omega_2}{2\Delta_2}\iproj{e}{g}), \label{eq:def-C2e}
\end{align}
and $\gamma_{j\xi}$ $(j = 1, 2, \xi = e, g)$ is the spontaneous decay rate associated with the $\ket{r_j} \rightarrow \ket{\xi}$ transition. These decay rates and their branching ratios (e.g. $\gamma_{1g}/\gamma_{1e}$) differ to a large extent among different physical systems depending on their specific energy structures.
%It is difficult to have a general solution applicable to an arbitrary system.
Here we first consider the simplest case where $\gamma_{1g} = \gamma_{1e} = \gamma_{2g} = \gamma_{2e}$ ($\equiv \gamma$) holds. In addition we assume that $\Delta_1 = \Delta_2$ ($\equiv \Delta$) and $\Omega_1 = \Omega_2$ ($\equiv \Omega$) hence $\chi = 0$ is satisfied. \ref{fig:avg_fid_atomic_decay} shows numerical calculations of the average gate fidelity using this setting for different values of $\Delta$ and $\delta$.

If the contributions from the terms proportional to the photon annihilation operator $a$ in the collapse operators (\ref{eq:def-C1g})-~(\ref{eq:def-C2e}) are negligible due to the small photon excitation, the effective spontaneous decay rate $\gamma_{\rm{eff}}$ is given by
\begin{align}
 \gamma_{\rm{eff}} \approx \gamma \frac{\Omega^2}{4\Delta^2}. \label{eq:gamma_eff}
\end{align}
Therefore the probability that a spontaneous emission occurs during the gate time is
\begin{align}
 P_{\mr{spont}} \sim \gamma_{\rm{eff}}t_{\rm{gate}} &= \frac{\pi\gamma\delta}{8g^2} = \frac{\pi\sqrt{m}\gamma\Omega}{4g\Delta}.
\end{align}
This probability needs to be sufficiently small in order for the atomic decay to be negligible for the gate fidelity. In that regard one can see that increasing $\delta$ is rather detrimental than beneficial, which can be also seen in \ref{fig:avg_fid_atomic_decay}.

\begin{figure}[t]
 \begin{center}
  \includegraphics[width=0.8\linewidth]{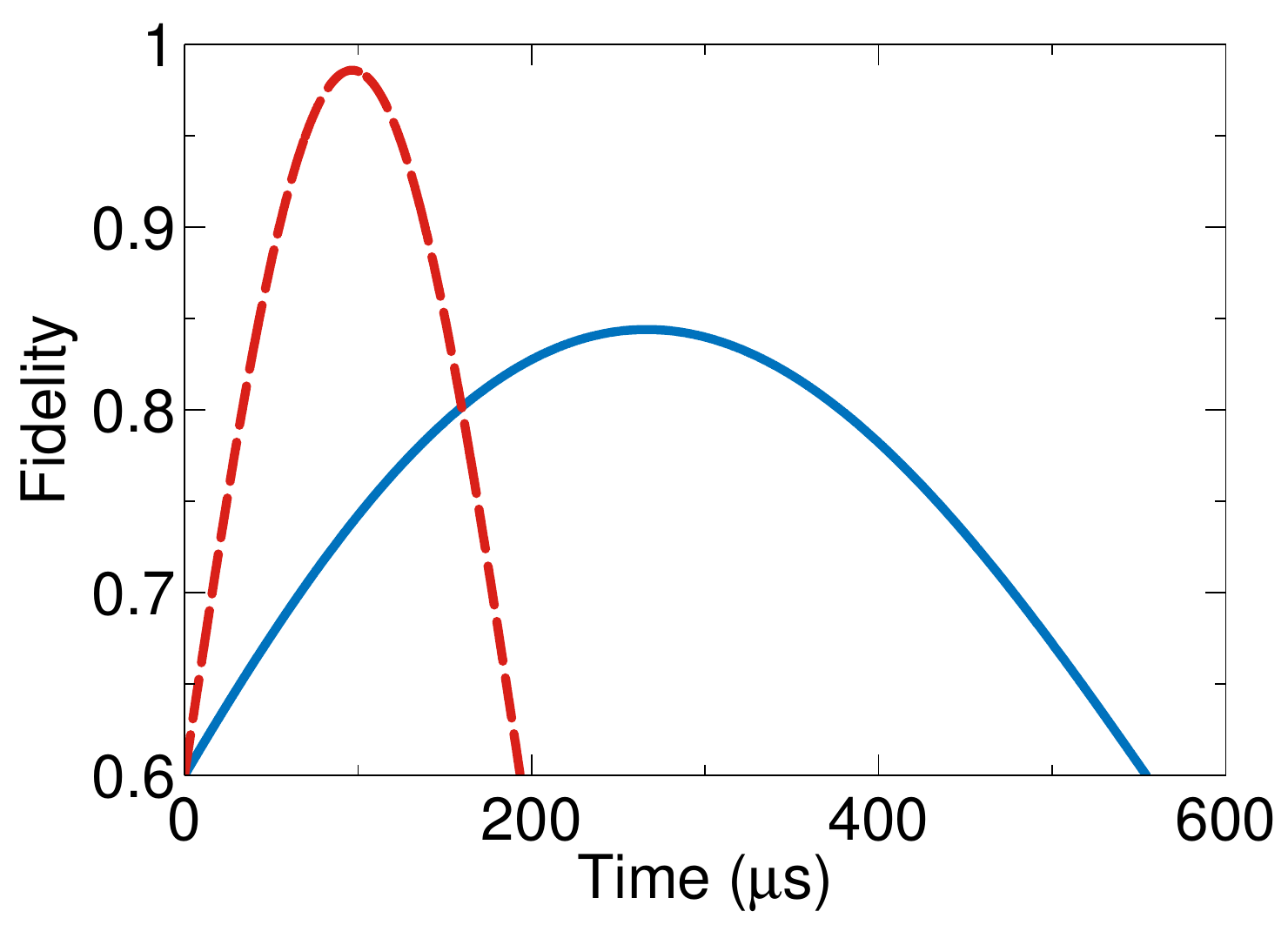}
  \caption{Average gate fidelities as a function of time for two sets of parameters in Table \ref{tab:Rb87-params}. Set 1 (blue solid) and set 2 (red dashed).}
  \label{fig:Rb87-avg-gate-fidelity}
 \end{center}
\end{figure}

In the general case where the cavity and atomic decays are both simultaneously present in the system, one needs to carefully choose experimental parameters depending on the relative magnitudes of $\kappa$ and $\gamma$ to minimize their total effect on the gate performance. As an example of such practical systems, we now consider single neutral ${}^{87}\rm{Rb}$ atoms coupled to a high finesse optical cavity. Note that this particular system was theoretically and experimentally studied in \cite{grimsmo2013cavity} and \cite{baden2014realization} respectively. See Appendix \ref{sec:model-87mrrb-atoms} for the details of modeling this system. First we assume a state-of-the-art conventional Fabry-Perot cavity with a length of $50\,\mr{\mu m}$ and finesse of $10^6$. This results in $g/2\pi = 60$ MHz and $\kappa/2\pi = 1.5$ MHz for a single $\Rb$ atom.
With the other parameters listed in set 1 of Table \ref{tab:Rb87-params}, we obtain $\geff/2\pi = 123\,\mr{kHz}$ and $\chi/2\pi = -240\,\mr{kHz}$,  and the maximum average gate fidelity of $\bar{F} = 0.844$ at $t = 260\,\mr{\mu s}$ (see~\ref{fig:Rb87-avg-gate-fidelity} ).
In order to further improve the gate fidelity, one can use an ultra-high Q  micro-sphere/toroidal cavity where a greater atom-cavity coupling is expected. Here we assume  $(g, \kappa)/2\pi = (200, 0.1)$ MHz \cite{grimsmo2013cavity} together with the parameters listed in set 2 of Table \ref{tab:Rb87-params}. Then we obtain $\geff/2\pi = 204\,\mr{kHz}$, $\chi/2\pi = -331\,\mr{kHz}$ and $\bar{F} = 0.986$ at $t = 98\,\mr{\mu s}$ (\ref{fig:Rb87-avg-gate-fidelity}).
\begin{table}[t]
 \begin{tabular}{| C{0.9cm} | C{0.9cm} | C{0.9cm} | C{1cm} | C{1cm} | C{0.8cm} | C{0.8cm} | C{0.8cm} |}
  \hline
 Set  & $g$ & $\kappa$ & $\Delta_1$ & $\Delta_2$ & $\Omega_1$ & $\Omega_2$ & $\delta$\\
\hline
1 & 60 & 1.5 & 10000 & 3980 & -50 & -17.6 & 19.6 \\
\hline
2 & 200 & 0.1 & 20000 & 13977 & -50 & -33.9 & 16.3 \\
\hline
 \end{tabular}
 \caption{Parameter sets used in the MS-gate simulations for $\Rb$ atoms. All the numbers are in units of $2\pi$ MHz.}
 \label{tab:Rb87-params}
\end{table}

\section{Conclusion}

In this article we have proposed a novel entangling quantum logic gate for general cavity QED systems. This scheme is inspired by the M{\o}lmer-S{\o}rensen gate in ion traps whose well-established physics enables us to have a clear picture of the basic gate dynamics and the effects of the imperfections. We have analyzed and evaluated possible adverse effects caused by the cavity-induced ac Stark shift, cavity field decay and atomic spontaneous emissions. Using the $\Rb$ system as a practical example, we have demonstrated that a high-fidelity gate operation is possible with sufficiently large atom-cavity coupling. However we believe that our scheme is applicable to a broader class of physical systems, not limited to atomic cavity QED, due to the fact that cavity interactions are ubiquitous in many different physical systems. In particular, nitrogen-vacancy centers in diamond and semiconductor quantum dots may serve as good qubit candidates in solid state, due to the rapidly improving optical control of their electronic quantum states \cite{gao2015coherent}.

In ion traps, there have been various proposals for extension of the MS-gate and entangling operations using a state-dependent force in general \cite{garcia2003speed,duan2004scaling,palmero2016fast}. Using the framework that we presented in this article, potentially these schemes can be also translated to cavity QED and used to further improve the gate fidelity.        
Furthermore, we would like to point out that the realization of the AJC Hamiltonian in cavity QED means that not only the MS-gate but many techniques developed in the context of ion trapping are transferable to optical cavity QED systems. Such examples can include engineered spin-spin interactions \cite{porras2004effective} and quantum state preparation of the bosonic modes \cite{kienzler2016observation,meekhof1996generation}. Moreover, the open-system nature of cavity QED could add another interesting aspect to such realizations through the external driving and extraction of the optical field.     

\begin{acknowledgments}
We gratefully acknowledge support from EPSRC through the UK Quantum Technology Hub: NQIT - Networked Quantum Information Technologies (EP/M013243/1 and
EP/J003670/1). H.T. and P.N.S thank Diego Porras for helpful discussions.  
\end{acknowledgments}

\appendix
\section{Perturbative expansion of the propagator in an interaction picture}
\label{sec:pert-expans-prop}
Starting from Hamiltonian (\ref{eq:Heff-prime2}) and moving on to an interaction picture defined by a unitary operator
\begin{align}
 U_\mr{I}(t) = \e^{-i\delta t a^{\dag}a} \label{eq:U_I},
\end{align}
we get a new system Hamiltonian
\begin{equation}
H_I = \chi a^{\dag}a S_z + \delta a a^\dag + g(a + a^\dag)S_x. \label{eq:HI}
\end{equation}
We define a new Hamiltonian
\begin{align}
H_{\mr{MS}}' &= \delta a^\dag a + g_\mr{eff}(a + a^\dag)S_x \label{eq:H-MS}
\end{align}
such that $H _I = H_{\mr{AS}} + H_{\mr{MS}}'$. 
It can be easily shown that the following states form a complete set of the energy eigenstates of $H_\mr{MS}'$ for $N = 2$. 
\begin{align}
&\ket{S=1, S_x=1}\ket{n, \alpha}, \label{eq:eigenstate-Sx1}\\
&\ket{S=1, S_x=0}\ket{n}, \label{eq:eigenstate-Sx0}\\
&\ket{S=1, S_x=-1}\ket{n, -\alpha}, \label{eq:eigenstate-Sx-1}\\
&\ket{S=0, S_x=0}\ket{n} \label{eq:eigenstate_S0-Sx0}
\end{align}
where $\ket{n, \alpha} = D(\alpha)\ket{n}$ and $D(\alpha)$ is the displacement operator with an amplitude $\alpha = -g/\delta$. Their energy eigenvalues are a function of $S_x$ and $n$ but not of $S$ and are given by
\begin{align}
 E_{jn} = \delta(n-(j\alpha)^2), \label{eq:Ejn}
\end{align}
for $\ket{S_x=j}\ket{n, j\alpha}$ $(j = 0, \pm 1)$.
%Therefore the states (\ref{eq:eigenstate-Sx0}) and (\ref{eq:eigenstate_S0-Sx0}) are energetically degenerate. 

Now we further move on to a second interaction picture with
\begin{align}
 U_{\mr{II}}(t) = \e^{iH'_{\mr{MS}} t}, \label{eq:U-II}
\end{align}
that leads to a new system Hamiltonian
\begin{equation}
H_\mr{II}(t) = \e^{iH'_{\mr{MS}}t} H_{\mr{AS}} \e^{-iH'_{\mr{MS}'}t}. \label{eq:HII}
\end{equation}
A wave function in this second interaction picture $\ket{\Psi_{\mr{II}}(t)}$ is related to the one in the original picture $\ket{\Psi(t)}$ by
\begin{align}
 \ket{\Psi_\mr{II}(t)} = U_{\mr{II}}(t)U_{\mr{I}}(t)\ket{\Psi(t)}. \label{eq:Psi-PsiII}
\end{align} 
On the other hand, the time evolution of $\ket{\Psi_\mr{II}(t)}$ is described by a propagator $V_{\mr{II}}(t)$.
\begin{align}
 \ket{\Psi_\mr{II}(t)} &= V_{\mr{II}}(t)\ket{\Psi(0)} \label{eq:time-evolution-PsiII}\\
 V_{\mr{II}}(t) &= 1- i\int_0^tH_\mr{II}(t')\,dt'\nonumber\\
 &+(-i)^2\int_0^t\!\int_0^{t'}H_\mr{II}(t')H_\mr{II}(t'')\,dt'dt''+\hdots \label{eq:V-Dyson-series}
\end{align}
where we have used a Dyson series expansion for $V_{\mr{II}}(t)$ as $H_{\mr{II}}(t)$ is not commutative at different times. Note also that the initial state $\ket{\Psi(0)}$ is identical among all the pictures. From (\ref{eq:Psi-PsiII}) and (\ref{eq:time-evolution-PsiII}), we get
\begin{align}
 \ket{\Psi(t)} = U_{\mr{I}}^\dag(t)U_{\mr{II}}^\dag(t)V_{\mr{II}}(t)\ket{\Psi(0)}. \label{eq:time-evolution-Psi}
\end{align}
From (\ref{eq:time-evolution-V})
\begin{align}
 V(t) = U_{\mr{I}}^\dag(t)U_{\mr{II}}^\dag(t)V_{\mr{II}}(t).
\end{align}

Since $H_{\mr{II}}(t)\ket{S=0, S_x=0} = 0$ and $[H_{\mr{II}}(t), S^2] = 0$, non-zero matrix elements of $H_{\mr{II}}(t)$ are limited in the $S=1$ manifold.
In other words $\ket{S=0, S_x=0}\ket{n}$ with an arbitrary photon number $n$ is an eigenstate of the perturbative Hamiltonian $H_\mr{AS}$ with an eigenvalue of 0. Therefore this state is not affected by the perturbation and the gate works for this state even in the presence of $H_\mr{AS}$.

From now on we only consider the matrix elements of $H_{\mr{II}}(t)$ in the $S=1$ manifold and we use the following abbreviation for the state notation:
\begin{align}
 \dket{j, n} = \ket{S=1, S_x=j}\ket{n, j\alpha}, \quad j = -1, 0, 1. \label{eq:def-jn-ket} 
\end{align}
The matrix elements of $H_{\mr{II}}(t)$ are calculated as follows:
\begin{align}
 \dbra{1, m} H_{\mr{II}}(t) \dket{1, n} &= 0, \label{eq:HII-matirx1}\\
 \dbra{-1, m} H_{\mr{II}}(t) \dket{1, n} &= 0, \label{eq:HII-matrix2}\\
 \dbra{0, m} H_{\mr{II}}(t) \dket{1, n} &= \frac{m\chi}{\sqrt{2}}\e^{i (E_{0m}-E_{1n})t}\bra{m}D(\alpha)\ket{n} \nonumber\\
 &=\chi \Lambda_{mn}(t;\alpha), \label{eq:HII-matrix3}\\
 \dbra{0, m} H_{\mr{II}}(t) \dket{-1, n} &= \frac{m\chi}{\sqrt{2}}\e^{i (E_{0m}-E_{1n})t}\bra{m}D(-\alpha)\ket{n} \nonumber\\
 &=\chi \Lambda_{mn}(t;-\alpha). \label{eq:HII-matrix4}
\end{align}
Here we have defined
\begin{align}
 \Lambda_{mn}(t;\alpha) = \frac{m}{\sqrt{2}}\e^{i (E_{0m}-E_{1n})t}\bra{m}D(\alpha)\ket{n}. \label{eq:def-Lambda}
\end{align}
With these matrix elements $H_\mr{II}(t)$ can be expressed as 
\begin{align}
 H_\mr{II}(t) &= \chi \sum_{m, n}\left(\Lambda_{mn}(t;\alpha)\dket{0,m}\dbra{1,n} \right.\nonumber\\
 &+ \left.\Lambda_{mn}(t;-\alpha)\dket{0,m}\dbra{-1,n}\right) + \hc \label{eq:HII-expansion} 
\end{align}
Likewise
\begin{align}
 &H_\mr{II}(t)H_\mr{II}(t') \nonumber\\ 
&= \chi^2\sum_{l,m,n}\left(\Lambda_{ml}(t;\alpha)\Lambda_{nl}^\ast(t';\alpha)+\Lambda_{ml}(t;-\alpha)\Lambda_{nl}^\ast(t';-\alpha)\right) \nonumber\\
 &\times\dket{0, m}\dbra{0, n} \nonumber\\
 &+\chi^2\sum_{j,j'=\pm 1}\sum_{l,m,n}\Lambda_{lm}(t;j\alpha)\Lambda_{ln}^\ast(t';j'\alpha)\dket{j, m}\dbra{j', n}. \label{eq:HII2-expansion}
\end{align}
Higher order integrands in (\ref{eq:V-Dyson-series}) can be as well calculated straightforwardly.

\section{A model for $\Rb$ atoms coupled to an optical cavity}
\label{sec:model-87mrrb-atoms}

\begin{figure}[h]
 \centering
 \includegraphics[width=0.7\linewidth]{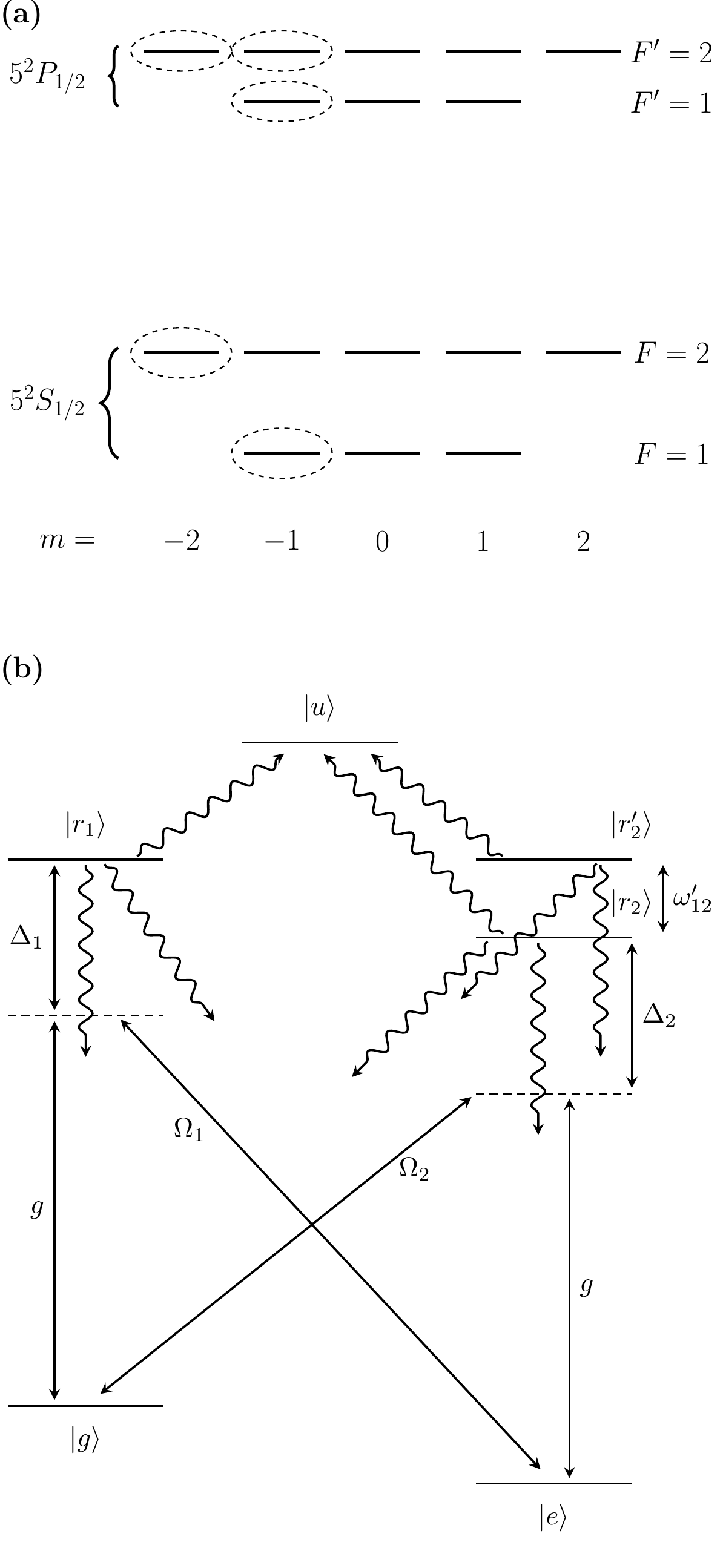}
 \caption{(a) Relevant energy levels for the D1 transition of ${}^{87}\mr{Rb}$ consisting of the $5^2S_{1/2}$ and $5^2P_{1/2}$ manifolds. The ones enclosed by the dashed ellipses are those explicitly included in the effective model. (b) The effective energy scheme employed in the calculation. The wavy arrows indicate the spontaneous decays from the excited states. $\ket{g} = \ket{F=2, m_F=-2}$, $\ket{e} = \ket{F=1, m_F=-1}$, $\ket{r_1} = \ket{F'=2, m_F=-2}$, $\ket{r_2} = \ket{F'=2, m_F=-1}$ and $\ket{r'_2} = \ket{F'=1, m_F=-1}$.}
 \label{fig:Rb87-level-scheme}
\end{figure}

We employ the $D_1$ transition between $5^2S_{1/2}$ and $5^2P_{1/2}$ states of single $\Rb$ atoms.  Among the eight Zeeman sublevels in the $5^2S_{1/2}$ manifold, we pick up $\ket{F=2, m_F=-2}$ and $\ket{F=1, m_F=-1}$ as our qubit states $\ket{g}$ and $\ket{e}$ respectively. These qubit states can be coupled to each other via the upper $5^2P_{1/2}$ states. Due to the relatively small hyperfine splitting in the $5^2P_{1/2}$ manifold (=812 MHz), we take not only $\ket{F'=1, m_F=-1}$ (=$\ket{r_2}$) but $\ket{F'=2, m_F=-1}$ (=$\ket{r'_2}$) into account as a mediating upper level for one of the cavity-assisted Raman transitions (see \ref{fig:Rb87-level-scheme}). The expressions for the effective parameters are obtained as follows \cite{grimsmo2013cavity}:
\begin{align}
 \geff^{(1)} &= \frac{g\Omega_1}{\sqrt{6}\Delta_1}, \label{eq:geff1-Rb87}\\
 \geff^{(2)} &= \frac{g\Omega_2}{2\sqrt{6}}\left(\frac{1}{\Delta_2}+\frac{1}{\Delta_2+\omega'_{12}}\right), \label{eq:geff2-Rb87} \\
 \chi &= g^2\left(\frac{1}{4(\Delta_2+\omega'_{12})} + \frac{1}{12\Delta_2}-\frac{1}{3\Delta_1} \right). 
\end{align}
Here $\omega'_{12}/2\pi = 812\,\mr{MHz}$ and $\geff^{(1)}$ and $\geff^{(2)}$ are the effective coupling strengths of the Raman transitions corresponding to the JC and AJC terms respectively. Note that the above expressions are modified from (\ref{eq:chi}) and (\ref{eq:geff}) due to the Clebsch-Gordan coefficients and the off-resonant coupling to $\ket{r'_2}$. The condition $\geff^{(1)} = \geff^{(2)}$ imposes a constraint for $\Omega_1$ and $\Omega_2$.

Each upper state can decay to the ground qubit states at a rate proportional to the Clebsch-Gordan coefficient for the relevant transition. In addition they can also decay to other Zeeman sublevels in the $5^2S_{1/2}$ manifold which are not shown in \ref{fig:Rb87-level-scheme}, effectively bringing the system out of the qubit subspace. In order to incorporate such decays into the model, we introduce a virtual auxiliary level $\ket{u}$. For example the decay rates to $\ket{u}$ from $\ket{r_1}$ is equal to the total sum of all the decay rates from $\ket{r_1}$ to the ground states except for the ones for $\ket{r_1} \rightarrow \ket{g}$ and $\ket{r_1} \rightarrow \ket{e}$. The same applies to the decays from $\ket{r_2}$ and $\ket{r'_2}$ to $\ket{u}$. In this way the decays to $\ket{u}$ embody all the decays to the outside of the qubit subspace. Note that in the real system it is possible for the atomic population outside the quibt subspace to be pumped back to the subspace again. Here we ignore such processes and the population in $\ket{u}$ only accumulates in the simulation.

In the end we have nine different Lindblad operators per atom in the form of (\ref{eq:Lindbladian}) with the following collapse operators (here 2$\gamma = 2\pi \cdot 5.75\,\mr{MHz}$):
\begin{align}
 C_{1g}^{(i)} &= \sqrt{\frac{\gamma}{3}}\left(\frac{g}{\sqrt{3}\Delta_1}a\iproj{g}{g} + \frac{\Omega_1}{2\sqrt{2}\Delta_1}\iproj{g}{e}\right), \label{eq:def-C1g-Rb} \\
 C_{1e}^{(i)} &= \sqrt{\frac{\gamma}{2}}\left(\frac{g}{\sqrt{3}\Delta_1}a\iproj{e}{g} + \frac{\Omega_1}{2\sqrt{2}\Delta_1}\iproj{e}{e}\right), \label{eq:def-C1e-Rb} \\
 C_{2g}^{(i)} &= \sqrt{\frac{\gamma}{2}}\left(\frac{g}{2\sqrt{3}\Delta_2}a\iproj{g}{e} + \frac{\Omega_2}{2\sqrt{2}\Delta_2}\iproj{g}{g}\right), \label{eq:def-C2g-Rb} \\
 C_{2e}^{(i)} &= \frac{1}{2}\sqrt{\frac{\gamma}{3}}\left(\frac{g}{2\sqrt{3}\Delta_2}a\iproj{e}{e} + \frac{\Omega_2}{2\sqrt{2}\Delta_2}\iproj{e}{g}\right), \label{eq:def-C2e-Rb}
\end{align}
\begin{align}
 C_{2'g}^{(i)} &= \sqrt{\frac{\gamma}{6}}\left(\frac{g}{2(\Delta_2+\omega'_{12})}a\iproj{g}{e} \right. \nonumber\\  &\qquad\qquad\qquad\left.+ \frac{\Omega_2}{2\sqrt{6}(\Delta_2+\omega'_{12})}\iproj{g}{g}\right), \label{eq:def-C2'g-Rb} \\
 C_{2'e}^{(i)} &= \frac{\sqrt{\gamma}}{2}\left(\frac{g}{2(\Delta_2+\omega'_{12})}a\iproj{e}{e} \right. \nonumber \\ &\qquad\qquad\qquad\left.+ \frac{\Omega_2}{2\sqrt{6}(\Delta_2+\omega'_{12})}\iproj{e}{g}\right), \label{eq:def-C2'g-Rb} \\
 C_{1u}^{(i)} &= \sqrt{\frac{5\gamma}{6}}\left(\frac{g}{2\sqrt{3}\Delta_1}a\iproj{u}{g} + \frac{\Omega_1}{2\sqrt{2}\Delta_1}\iproj{u}{e}\right), \label{eq:def-C1u-Rb} \\
 C_{2u}^{(i)} &= \sqrt{\frac{17\gamma}{12}}\left(\frac{g}{2\sqrt{3}\Delta_2}a\iproj{u}{e} + \frac{\Omega_2}{2\sqrt{2}\Delta_2}\iproj{u}{g}\right), \label{eq:def-C2u-Rb} \\
 C_{2'u}^{(i)} &= \sqrt{\frac{19\gamma}{12}}\left(\frac{g}{2(\Delta_2+\omega'_{12})}a\iproj{u}{e} \right.\nonumber \\ &\qquad\qquad\qquad\left.+ \frac{\Omega_2}{2\sqrt{6}(\Delta_2+\omega'_{12})}\iproj{u}{g}\right). \label{eq:def-C2'u-Rb}
\end{align}

\bibliographystyle{apsrev4-1}
\bibliography{cavity_MS}

\end{document}